\documentclass{article}

\usepackage{arxiv}

\usepackage[utf8]{inputenc}
\usepackage[T1]{fontenc}   
\usepackage{hyperref}       
\usepackage{url}            
\usepackage{booktabs}      
\usepackage{amsfonts}       
\usepackage{nicefrac}       
\usepackage{microtype}      
\usepackage{amsmath}
\usepackage{cleveref}       
\usepackage{graphicx}
\usepackage{natbib}
\usepackage{doi}
\usepackage{float}
\usepackage{graphicx}
\usepackage{subcaption}
\usepackage{adjustbox}
\usepackage{algorithm}
\usepackage{algpseudocode}
\usepackage{makecell}

\DeclareMathOperator*{\argmax}{arg\,max}

\title{Longitudinal market structure detection using a dynamic modularity-spectral algorithm}

\newif\ifuniqueAffiliation

\uniqueAffiliationtrue

\ifuniqueAffiliation 
\author{Philipp Wirth
			\\
	Institute of Finance and Technology\\
	University College London\\
	London, WC1E 6BT\\
	\texttt{philipp.wirth@ucl.ac.uk} \\
	\And
	Francesca Medda \\
	Institute of Finance and Technology\\
	University College London\\
	London, WC1E 6BT\\
	\texttt{f.medda@ucl.ac.uk} \\
	\AND
	Thomas Schröder \\
	Institute of Finance and Technology\\
	University College London\\
	London, WC1E 6BT\\
	\texttt{t.schroeder@ucl.ac.uk} \\
	}

\renewcommand{\shorttitle}{DynMSA}

\hypersetup{
pdftitle={Longitudinal market structure detection using a dynamic modularity-spectral algorithm},
pdfsubject={q-fin.NC, q-fin.PM, cs.CE, cs.LG},
pdfauthor={Philipp Wirth},
pdfkeywords={Portfolio Management, Equity Clustering, Graph-Based Clustering, Modularity Optimisation, Spectral Clustering, Diversification, Asset Allocation, Risk Management},
}

\begin{document}
\maketitle

\begin{abstract}
In this paper, we introduce the Dynamic Modularity-Spectral Algorithm (DynMSA), a novel approach to identify clusters of stocks with high intra-cluster correlations and low inter-cluster correlations by combining Random Matrix Theory with modularity optimisation and spectral clustering.
The primary objective is to uncover hidden market structures and find diversifiers based on return correlations, thereby achieving a more effective risk-reducing portfolio allocation.
We applied DynMSA to constituents of the S\&P 500 and compared the results to sector- and market-based benchmarks. Besides the conception of this algorithm, our contributions further include implementing a sector-based calibration for modularity optimisation and a correlation-based distance function for spectral clustering.
Testing revealed that DynMSA outperforms baseline models in intra- and inter-cluster correlation differences, particularly over medium-term correlation look-backs. It also identifies stable clusters and detects regime changes due to exogenous shocks, such as the COVID-19 pandemic. Portfolios constructed using our clusters showed higher Sortino and Sharpe ratios, lower downside volatility, reduced maximum drawdown and higher annualised returns compared to an equally weighted market benchmark.
\end{abstract}


\section{Introduction}
In the world of financial markets, cluster analysis has long been a viable tool for understanding the dynamics between assets, especially considering risk management and portfolio diversification. It is long known and accepted that assets are interconnected over sectors, countries and asset classes. While arguably the first formalisation of this assumption dates back to Howard Markowitz and his Modern Portfolio Theory (MPT) from 1952 \citep{markowitz1952portofolio}, events like the Great Depression in 1929 already exhibited the behaviour that stocks move in the same direction despite being seemingly different. 

Later on, Mantegna formulated the idea of a hierarchical market structure in his seminal paper \citep{mantegna1999hierarchical}. To find clusters of stocks that behave similarly, Mantegna used asset price correlations and mapped them to a minimum spanning tree (MST). In recent years, researchers and practitioners have explored various other ways to cluster assets such as hierarchical clustering based on Planar Maximally Filtered Graphs \citep{pozzi2013spread}, community detection using modularity optimisation \citep{macmahon2015community} and random matrix theory \citep{plerou2000random}. Many of these approaches aim to understand the interconnectivity of stocks and other assets for risk management purposes. 

In an ideal portfolio, investors want to diversify their holdings in a way to optimise their risk exposure based on personal risk levels and return expectations. This is not a novel approach. The benefits of diversification have been described by \citet{markowitz1952portofolio} and \citet{roy1952safety} over 70 years ago. Clustering methods can help manage stock risk by grouping assets with similar behaviours, making it easier to identify and mitigate sector-specific or systematic risks. However, traditional clustering methods like k-means require predefined parameters, which are impractical due to changing market dynamics and evolving stock relationships.

By applying graph theory, some limitations of traditional clustering algorithms can be circumvented. A correlation matrix of stock returns can be represented as a fully connected graph, where the nodes are stocks and the edges represent correlations. Graph clustering algorithms, such as community detection using modularity optimisation \citep{newman2004finding}, leverage the interconnected structure of stocks to identify inherent and naturally occurring clusters without predefined numbers of clusters and stocks per cluster. These methods provide a more dynamic and nuanced way of clustering stocks.

The goal of this paper is to develop and evaluate a novel, longitudinal clustering framework for equities, leveraging the core benefits of graph-based clustering. Specifically, this research will focus on the following:

\begin{itemize}
\item \textit{Longitudinal clustering using community detection and spectral clustering:}
Constructing a method for dynamically clustering equities over time, utilising two core principles of graph clustering methods - namely, the capability to inherently model the interconnectedness of financial markets and the non-requirement of defining the number of clusters a priori, as needed with other clustering methods.
\end{itemize}

To achieve these objectives, we have developed a novel, graph-based clustering algorithm called the Dynamic Modularity-Spectral Algorithm (DynMSA), combining Random Matrix Theory (RMT) with modularity optimisation and spectral clustering. The hypotheses tested in this research include:

\begin{itemize}
    \item \textit{Intra- and inter-cluster correlation differences:} DynMSA is expected to yield clusters with significantly higher intra-cluster correlations and lower inter-cluster correlations compared to sector-based clusters.
    \item \textit{Combined approach effectiveness:} A combined approach of modularity optimisation and spectral clustering is anticipated to outperform using modularity optimisation alone.
    \item \textit{Cluster stability and regime changes:} The method should identify stable clusters over time while also detecting regime changes due to exogenous shocks such as the COVID-19 pandemic.
    \item \textit{Hidden market strucutre}: The currently used sector classification is not perfectly accurate. Stocks are attributed to sectors that they empirically do not belong to. DynMSA is expected to find these hidden market structures.
    \item \textit{Portfolio construction benefits:} We assume that DynMSA can find stocks that are necessarily attributable to certain clusters and that such stocks are better diversifiers. Therefore, portfolios constructed using our clustering method are expected to exhibit superior risk-adjusted performance metrics, such as higher Sortino and Sharpe ratios, compared to an equally weighted market benchmark. 
\end{itemize}

We add to the existing literature by developing this novel ensemble framework and analyse the clusters on a monthly basis over several years, including testing if exogenous shocks like COVID-19 lead to shifts in correlation patterns and therefore different stock relationships.

The remainder of this work is structured as follows: We start by giving a brief overview of related work in the fields of clustering in financial markets and graph clustering. Then we introduce DynMSA before explaining the concepts and novelties used in building our algorithm, including community detection, spectral clustering and the stock selection process. Lastly, we will focus on analysing the outcomes of the applying DynMSA to a data set containing US equities.

\section{Diversification in equity markets}

\subsection{Importance}
Building on the understanding that equities are highly connected and share similar behaviours during different market regimes, diversification is a fundamental strategy for managing risk and improving portfolio returns. The importance of diversification has been extensively studied. 

The foundational work of \citet{markowitz1952portofolio} introduced MPT which emphasises as one of its corner stones the benefits of diversifying between various assets to reduce portfolio risks. Despite that knowledge, \citet{goetzmann2008equity} found that the individual US investors hold under-diversified portfolio which results in a welfare loss for these investors. 

The behaviour of equities during extreme market conditions has been a key area of study. \citet{longin2001extreme} used extreme value theory to model the tails of multivariate return distributions for testing the hypothesis, if stocks are higher correlated during more volatile periods. They found that correlations indeed increase in bear, but not in bull markets. This finding was later supported by \citet{ang2002asymmetric} who showed that correlations for US equities are much higher during periods of downside stress, particularly for extreme downside moves. This asymmetry poses challenges for diversification, as correlation between stocks increases exactly when diversification is needed the most. Empirically this could be observed only a few years later during the Great Financial Crisis.

\citet{chua2009myth} argued therefore, that upside diversification is actually not desired as this could limit the returns of a portfolio. They proposed a full-scale optimisation method to produce portfolios where stocks are highly correlated in bull markets while being less correlated in bear markets, relative to the classic mean-variance optimisation. \citet{kinlaw2021myth} addressed this topic again a few years later, looking at asset pairs to optimise portfolio performance. 

More recent studies have further highlighted the interconnectedness in equity markets, particularly during periods of higher market stress. \citet{akhtaruzzaman2021financial} found that financial contagion during the COVID-19 pandemic led to significant increases in conditional correlations between stock returns of firms in China and G7 countries, particularly among financial firms. \citet{jalloul2023equity} analysed and found higher connectedness of stocks during geopolitical crises and exogenous shocks, particularly terrorist attacks. 

\citet{feldman2017contagious} used sentiment analysis to partially explain why return correlations between different markets increase during financial crises, adding a behavioural dimension to the understanding of diversification. \citet{diebold2009measuring} developed measures of return and volatility spillovers, finding divergent behaviours in their dynamics: while return spillovers exhibit an increasing trend, volatility spillovers show no such trend. However, in times of crises, volatility spillovers spike significantly, supporting the view that in market downturns diversification becomes highly relevant to smoothing the overall risk in these periods.

\citet{yen2021understanding} attacked the research about market interdepencies from a slightly different angle and introduced a model based on topological data analysis. They show complex geometric structures in financial networks which again highlight the high level of interconnectedness of financial markets.

\subsection{Methods and challenges}
Despite the clear evidence for the importance of diversification, achieving it remains challenging and the traditional method based on MPT has several drawbacks. Most dominantly:
\begin{itemize}
    \item \textbf{Return Prediction:} MPT requires the prediction of future returns, which is inherently uncertain and can lead to significant errors in portfolio construction \citep{markowitz1952portofolio}.
    \item \textbf{Covariance Matrix Estimation:} Estimating the covariance matrix accurately is challenging due to the noisy nature of financial data and can often lead to estimation errors, as extreme values in small sample sizes are disproportionately high considered \citep{ledoit2004honey}.
\end{itemize}

To counter these problems, various other methods for asset allocation and diversification were developed. This ranges from understanding different performance driving factors, such as those in the Fama-French models \citep{fama1993common, fama2015five} and investing along these factors, to improving MPT by including shrinkage methods to calculate the covariance matrix \citep{ledoit2004honey}.

Another strand of the literature focuses on leveraging the aforementioned asset correlations by understanding the underlying drivers, with sector-based influences being one of the potential forces behind these connections.

While \citet{bekaert2009international} argued that the rising importance of sectors as a diversification is less relevant than previous papers suggested when examining not only industries, but also country-style portfolios, other authors showed the influence sectors have on stock returns. 

\citet{heiberger2014stock} used a network approach to provide insights into the structural changes of equity networks during financial crises. He showed, that a network consisting of S\&P 500 stocks, which exhibited distinctive, partly sector-based clusters in normal times, tightens and becomes more centralised during periods of market stress, such as the dotcom bubble and the Great Financial Crisis. Analysing the dependencies of approximately 4,000 global stocks, \citet{raddant2021interconnectedness} highlighted the role of global sectoral factors in equity markets. Their findings indicated that sectors such as Energy, Materials, and Financials play a leading role in connecting markets, with the interconnectedness being volatile and influenced by changes in volatility and statistical interdependence across countries.

Diversifying along sectors however is not an objective task, as the classification of stocks in sectors and industries is typically done by rating agencies using proprietary methodologies \citep{phillips2016industry}. Additionally, as some companies operate in several business areas, the classification in one specific sector is not necessarily straightforward. Another study also found, that the sector influence on smaller stocks is not as high as it is on larger stocks \citep{chan2007industry}.

These difficulties and the subjectiveness with sector-based classification call for more objective solutions, such as risk-based models and graph-based models. Purely risk-based models such as risk parity \citep{qian2011risk} and hierarchical risk parity \citep{de2016building} require much less information than MPT, factor or sector models as they rely exclusively on historic returns to compute the necessary covariance matrix. These risk-based methods are more objective, particularly when used as graph-based approaches like hierarchical risk parity. 

Graph-based models in general have gained in popularity, as mentioned in the introduction. Some methods, such as the aforementioned minimum spanning trees \citep{mantegna1999hierarchical} and planar maximally filtered graphs \citep{tumminello2005tool} divide asset networks based on their hierarchical properties, while other methods such as community detection, focus on cluster similarities within a network. Modularity optimisation has been applied to equities in different ways (see, e.g., \citet{macmahon2015community}, \citet{kocheturov2014dynamics}, \citet{sim2021can}, \citet{millington2021stability}) and is capable of finding a completely data-driven market structure in equity networks.

We focus on this approach too and aim to improve it by utilising the Leiden algorithm instead of the typically used Louvain \citep{blondel2008fast} method. The Leiden algorithm offers faster and more accurate community detection with better guarantees of convergence to high modularity partitions \citep{traag2019louvain}. Additionally, we propose to combine this with spectral clustering to verify the connections identified by the original modularity optimisation. Spectral clustering, which leverages the eigenvalues and eigenvectors of the graph Laplacian \citep{von2007tutorial}, can provide an independent check to determine if the stocks identified as unconnected in the modularity optimisation are indeed unconnected. By integrating these advanced techniques, we aim to achieve a more detailed and accurate understanding of market structures, ultimately enhancing our ability to construct well-diversified portfolios.

\section{Correlation networks and optimisation objective }

\subsection{Correlation network}

Correlation as a measure to understand the risk of stocks and assets in general has been used extensively in practice and academia. This stems from the idea that a highly diversified portfolio is more stable and prevents from sudden drawdowns compared to a highly concentrated portfolio. We calculate the correlation of daily stock returns by using the Pearson correlation coefficient at the end of each month using daily returns over the past $n \in {3, 6, 12, 24} $ months as follows:

\begin{equation}
\rho_{i,j}^{n} = \frac{\sum_{d \in D(n)} (r_{i,d} - \overline{r_{i,n}})(r_{j,d} - \overline{r_{j,n}})}{\sqrt{\sum_{d \in D(n)} (r_{i,d} - \overline{r_{i,n}})^2 \sum_{d \in D(n)} (r_{j,d} - \overline{r_{j,n}})^2}}
\end{equation}

where $r_{i,d}$ and $r_{j,d}$ are the daily returns of stocks $i$ and $j$, $D(n)$ is the set of all trading days in the past $n$ months and $\overline{r_{i,n}}$ and $\overline{r_{j,n}}$ are their respective average daily returns over that period.

Depending on the chosen monthly periods, this leaves us with 61 to 82 correlation matrices which form the basis of our clustering algorithm. In our analysis, we will use the full length of correlation matrices for all tests except for the portfolio construction to make the portfolio key performance indicators better comparable between the different correlation look-backs. 

\begin{table}[H]
\centering
	\begin{tabular}{|c|c|c|c|c|c|}
	\hline
	n - months & 3 & 6 & 12 & 24 \\
	\hline
	m - correlation matrices &82&79&73&61\\
	\hline
	\end{tabular}
\caption{Number m of correlation matrices for n-months rolling correlation}
\label{tab: num_corr_matrices}
\end{table}

\subsubsection*{Intra- and inter-cluster correlations}
\vspace{1ex}

Given the objective of finding communities of stocks that are highly correlated within a community and low/uncorrelated with stocks in different communities, we define two measurement functions we will apply throughout the analysis of this and the next chapter. First, we compute the average intra-cluster correlation

\begin{equation}
\rho_{intra}=\frac{1}{C}\sum_{i=1}^{C}\left(\frac{1}{\frac{N_{i}(N_{i}-1)}{2}}\sum_{j=1}^{N_{i}}\sum_{k=j+1}^{N_{i}}\rho(s_{C,j},s_{C,k})\right)
\end{equation}

where $C$ is the number of clusters, $N_{i}$ is the number of stocks per cluster and $\rho(s_{C,j},s_{C,k})$ is the correlation between stocks $j$ and $k$ in cluster $C$. Furthermore we calculate the average inter-cluster correlation

\begin{equation}
\rho_{inter}=\frac{1}{\frac{C(C-1)}{2}}\sum_{C_{1}=1}^{C}\sum_{C_{2}=C_{1}+1}^{C}\left(\frac{1}{N_{C_{1}}N_{C_{2}}}\sum_{i=1}^{N_{C_{1}}}\sum_{j=1}^{N_{C_{2}}}\rho(s_{C_{1},i},s_{C_{2},j})\right)
\end{equation}

where $\rho(s_{C_1,i},s_{C_2,j})$ is the correlation of stock $i$ from cluster $C_{1}$ with stock $j$ from $C_{2}$.

The function $\rho_{intra}$ sums over all correlations for each cluster and calculates the average correlation for each cluster by considering only the unique correlation pairs, excluding the diagonal values as well as the upper triangle of the correlation matrix. Finally it calculates the average correlation over all clusters.

The function $\rho_{inter}$ follows a similar approach. The correlation between different clusters is calculated on per stock basis, meaning that we sum over the correlations of each individual stock pair from $C_{1}$ and $C_{2}$. As before, we exclude the diagonal values as well as the upper triangle from the correlation matrix.

\subsection{Objective for clustering}
As written above, the objective of this work is to find clusters of stocks that are highly correlated within a cluster, but un- or negatively correlated in between clusters. Mathematically we can formalise this as an optimisation problem

\begin{equation}
    \label{eq_opt_objective}
        Q_o = \max \left( \rho_{intra} - \rho_{inter} \right)
\end{equation}

Below we will present and apply graph clustering techniques that inherently maximise this objective through their algorithmic design as well as explicitly use it in a brute force method to select the best cluster results.

\section{Dynamic Modularity-Spectral Algorithm (DynMSA)} \label{sec:DynMSA}
To effectively analyse inter-temporal market structures, we propose a novel algorithm called Dynamic Modularity-Spectral Algorithm (DynMSA). This algorithm combines top-down modularity optimisation with a bottom-up spectral clustering algorithm. These methods are wrapped in a clustering pipeline that is built as follows

\begin{algorithm}[H]
\caption{Cluster Assignments and Additional Metrics}
\begin{algorithmic}[1]
\Require Daily stock returns, sector information
\Ensure Cluster assignments, additional metrics, and portfolio selection

\State \textbf{Compute} correlation matrix $C$ from daily stock returns

\State \textbf{Preprocess} correlation matrix $C$:
   \State \hspace{\algorithmicindent} (a) Apply Marchenko-Pastur-based cleaning to obtain cleaned matrix $C_{\text{cleaned}}$
   \State \hspace{\algorithmicindent} (b) Create threshold matrix from $C_{\text{cleaned}}$

\State \textbf{Convert} threshold matrix into network $G$

\State \textbf{Define} initial community membership:
   \For{each stock in $G$}
      \State Assign community based on sector classification
   \EndFor

\State \textbf{Run} Leiden algorithm on network $G$:
   \State \hspace{\algorithmicindent} (a) Use modularity as the quality function
   \State \hspace{\algorithmicindent} (b) Obtain initial clusters

\State \textbf{Refine} clusters with spectral clustering:
   \For{each cluster in initial clusters}
      \If{cluster size is less than $n$}
         \State Compute similarity matrix for stocks in cluster
         \State Run spectral clustering on similarity matrix
         \State Compute objective functions for both clustering results
         \If{spectral clustering result is better}
            \State Update cluster with spectral clustering result
         \Else
            \State Keep initial cluster result
         \EndIf
      \EndIf
   \EndFor

\State \textbf{Select} $x$ stocks for portfolio:
   \State \hspace{\algorithmicindent} (a) Select all stocks from small clusters $(C_i \leq n)$:
   \State \hspace{\algorithmicindent} (b) Select stocks from larger clusters proportionally to their cluster size
   \State \hspace{\algorithmicindent} (c) Combine selected stocks from small and large clusters
   
\State \textbf{Return} final clusters, additional metrics for further computations (e.g., portfolio construction) and selected portfolio

\end{algorithmic}
\end{algorithm}

DynMSA combines existing methods for community detection and cluster analysis, namely modularity optimisation and spectral clustering, in a new and innovative way by working in tandem to optimise the objective laid out in Equation \ref{eq_opt_objective}. Furthermore, we introduce several novel substeps in our algorithm. We use the Leiden algorithm instead of the most commonly used Louvain algorithm for the community detection and add a novel, sector-based ground configuration to our stock network. For spectral clustering, we introduce a similarity matrix based on the unweighted pair group method with arithmetic mean instead of the more commonly used Gaussian approach. Lastly, we introduce a correlation ranking mechanism to select stocks for our portfolios. In the coming sections we will explain each step of the algorithm in detail.

\section{Community detection in correlation networks}

When we think about a visual representation of correlations between different assets, we typically think of correlation matrices or heatmaps. However, we can also model correlations as a network, in which assets represent the nodes and correlations represent the edges. This depiction allows us to instantly see, how connected different entities on capital markets actually are. It also allows us to employ techniques not typically used in classical finance, namely techniques based on graph theory. One such technique is the division of nodes in a network into clusters or communities that are densely populated. In other words, communities that share the same characteristics, i.e., are highly correlated. For a general overview over different community detection techniques, we refer the reader to \citet{fortunato2010community}.

\subsection{Correlation matrix cleaning using the Marchenko-Pastur theorem}
Correlation matrices are a fundamental tool in the analysis of stock behaviour as they are capable of capturing inter-dependencies across markets and sectors. However, research has shown that such matrices often contain a mix of genuine information and random noise \citep{laloux1999noise}, which can be attributed to noisy covariance matrices. The intensity of this noise seems to be highly dependent on the size $n$ of the portfolio or investment universe and the length $t$ of the available daily returns \citep{pafka2003noisy}.

Random Matrix Theory on the one hand shows that a significant proportion of the eigenvalues for correlation matrices are similar in their universal properties to a Gaussian orthogonal ensemble of random matrices which indicates significant randomness in return correlation matrices \citep{plerou2002random}. On the other hand, it provides a robust framework for distinguishing between random noise and meaningful signals in complex systems by relying on RMT techniques to clean the correlation matrix \citep{bun2017cleaning}. One such technique is applying correlation matrix cleaning based on the Marchenko-Pastur theorem.

\subsubsection*{Marchenko-Pastur Theorem}
The Marchenko-Pastur theorem was introduced by Volodymyr Marchenko and Leonid Pastur in 1967 and describes the distribution of eigenvalues for random matrices, particularly the limit of the empirical distribution of eigenvalues of sample covariance matrices as the matrix size grows under specific conditions \citep{marchenko1967distribution}. It is widely considered a seminal contribution to Random Matrix Theory.

The theorem applies under following assumptions:

\begin{description}
    \item[Independence:] Let $X$ be a $m \times n$ matrix. Then the matrix values need to be independent identically distributed (i.i.d.) with $\mu=0$ and $\sigma^2<\infty$.
    \item[Aspect ratio condition:] The size of the matrix, represented by the number of rows ($m$) and the number of columns ($n$), should satisfy the condition that as $m, n \rightarrow \infty$, the ratio $m/n \rightarrow \lambda$ where $\lambda \in (0,+\infty)$. This ratio is often referred to as the aspect ratio of the matrix. 
\end{description}

In the case that $\lambda \in (0,1)$, the Marchenko-Pastur density $\rho(x)$ can be written as 
\begin{equation}
    \label{eq_mp_distribution}
        \rho(x)=\frac{1}{2\pi\sigma^2} \frac{\sqrt{(x-\lambda_-)(\lambda_+ -x)}}{\lambda x}1_{[\lambda_-,\lambda_+]}(x)
\end{equation}
where $\lambda_-$ and $\lambda_+$ are the lower and upper bounds of the density function and are given by 

\begin{equation}
    \label{eq_mp_upperlower}
        \lambda_-=\sigma^2(1-\sqrt{\lambda)}^2, \lambda_+=\sigma^2(1+\sqrt{\lambda)}^2 
\end{equation}

Additionally, $1_{[\lambda_-,\lambda_+]}(x)$ indicates that the function is $1$ if $x$ is in $[\lambda_-,\lambda_+]$ and $0$ otherwise.

The Marchenko-Pastur distribution provides a way to understand the behaviour of eigenvalues under random conditions. If the eigenvalues lie within the bounds defined by the distribution, they are considered as noise and can be filtered out to improve the overall structure and information content of the correlation matrix. Conversely, if the eigenvalues lie outside these bounds, they are considered meaningful and systematic information that should be kept for further computation. This raises the question of how to proceed with the eigenvalues that are considered noise. \citet{laloux1999noise} propose to replace these values with a constant, specifically the average of the noisy eigenvalues. Based on this approach, we implement the following steps. Given an initial correlation matrix $C$, we compute the eigenvalue decomposition so that

\begin{equation}
    \label{eq_mp_eigenvalue_decomp}
        C = VDV^T
\end{equation}

where $V$ is a matrix of eigenvectors and $D$ is a diagonal matrix containing the eigenvalues $\lambda_i$ of $C$.

We then modify the eigenvalues that fall within the bounds of the Marchenko-Pastur theorem by replacing them with the average eigenvalue $\overline{\lambda}$ of the noisy eigenvalues, while retaining the original eigenvalues for those outside the bounds, indicating significant relationships within the network. Mathematically, we handle these eigenvalues as follows:

\begin{equation}
    \label{eq_eigenvalue_handling}
    \tilde{\lambda}_i = 
    \begin{cases} 
        \lambda_i, & \text{if } \lambda_i > \lambda_+ \text{ or } \lambda_i < \lambda_- \\
        \overline{\lambda} = \frac{1}{n} \sum_{\lambda_i \in [\lambda_-, \lambda_+]} \lambda_i, & \text{if } \lambda_- \leq \lambda_i \leq \lambda_+
    \end{cases}          
\end{equation}

In the next step, we reconstruct the correlation matrix using the modified eigenvalues $\tilde{\lambda}$ and the original eigenvectors $V$, so that \ref{eq_mp_eigenvalue_decomp} becomes

\begin{equation}
    \label{eq_mp_eigenvalue_decomp2}
        C_{cleaned} = V\tilde{D}V^T
\end{equation}

where $\tilde{D}$ is the diagonal matrix of modified eigenvalues $\tilde{\lambda_i}$.

\subsection{Threshold Matrix}
$C_{cleaned}$ now constitutes a fully connected network where each edge weight is defined by the modified eigenvalues $\tilde{\lambda_i}$ as well as the original eigenvectors. While Leiden is capable of dealing with large, fully connected networks, the tight band of reconstructed matrix elements for a large data set can still make it difficult to find well-defined communities within the network. Introducing some sparsity to the network can be beneficial, as the clearer delineation of community bounds could help the algorithm to converge faster \citep{newman2004finding}. By applying a threshold $\theta$ to the matrix elements of $C_{cleaned}$, we can introduce some sparsity and retain only the most significant connections. As \citet{traag2019louvain} pointed out, their algorithm is particularly efficient when handling networks where edge weights signify strong relationships. This reduction in complexity therefore not only decreases the computational time, but also increases the accuracy of the community detection process.

The application of thresholding a correlation matrix has been used for defining asset graphs for over two decades (see, e.g., \citet{onnela2004clustering} \& \citet{heimo2009maximal}). In RMT, \citet{plerou2002random} set all values lower than or equal to the upper Marchenko-Pastur bound to $0$. The beauty and simplicity in thresholding lies in the objective, that only the most relevant connections should be considered. However, implementing a threshold matrix now comes with the challenge of defining the optimal threshold $\theta^*$. We propose an algorithmic way by implementing a brute-force method, where different $\theta$ are tested based on the optimisation objective that we defined in \ref{eq_opt_objective}. To account for $\theta$ in this function, we need to slightly adapt it as follows

\begin{equation}
    Q_o(\theta) = \max(\rho_{intra}(\theta)-\rho_{inter}(\theta))
\end{equation}

With this we can compute the best $\theta^*$ as

\begin{equation}
    \label{eq_adj_obj_theta}
        \theta^* = \argmax_{\theta \in [0,\lambda_+]} Q_o(\theta)
\end{equation}
 where $\theta \in [0,\lambda_+]$ gives the bounds of possible values for $\theta$.

 Using the threshold, we can then apply this value to $C_{cleaned}$ by setting all values above the threshold equal to $1$ and all values below the threshold equal to $0$. This results in a final matrix $T$, which can be derived as follows:

 \begin{equation}
     \label{eq_threshold_matrix}
        T = H(C_{cleaned}-\theta^* J)
 \end{equation}

where $H$ denotes the Heaviside step function defined by

\begin{equation}
    \label{eq_heaviside}
        H(x)=
        \begin{cases}
            0, &\text{if } x \leq 0
            \\
            1, &\text{if } x > 0
        \end{cases}
\end{equation}

$H$ is then element-wise applied to the matrix $C_{cleaned}-\theta^* J$ where $J$ is a matrix of the same dimensions as $C_{cleaned}$ with all entries set to $1$. This results in the binary matrix $T$ where each entry $T_{ij}$ is defined by

 \begin{equation}
     \label{eq_threshold_matrix_single_values}
        T_{ij} = H(C_{cleaned}-\theta^*)
 \end{equation}

Following this, $T_{ij}$ is set to $1$ if the corresponding entry in $C_{cleaned}$ exceeds the threshold $\theta^*$ and $0$ otherwise.

\subsection{Community detection methods}

\subsubsection*{Modularity}
\vspace{1ex}
After creating a cleaned threshold matrix and transforming the data into a graph, we need to find the communities of stocks with the most dense connections. One way to find such communities is by using an algorithm that maximises modularity. \citet{newman2004finding} first described modularity as a quality function for their algorithm aiming to find the best clusters in a graph. The general assumption is, that a random graph should not have a detectable cluster structure. Modularity can be written as follows

\begin{equation}
\label{eq_modularity}
	Q_m=\frac{1}{2m}\sum_{ij}(A_{ij}-\frac{k_ik_j}{2m})\delta(C_{i},C_{j})
\end{equation}

where $A_{ij}$ is the adjacency matrix, $\frac{k_ik_j}{2m}$ is the value of expected edges in the null model with $k_i$ being the weighted degree of node $i$, $m$ is the total number of edges in the graph and the Kronecker delta $\delta$ defining if two vertices $i$ and $j$ are in the same community. The null model is a randomised version of the original network, used to determine if the communities found from the clustering are actually different from random and therefore statistically significant. The null model is typically defined as a random graph with the same degree distribution as the original graph.

\subsubsection{Modularity optimisation - Louvain vs. Leiden}
\vspace{1ex}

One of the most commonly used algorithms for community detection based on modularity is the Louvain algorithm, developed by \citeauthor{blondel2008fast} in 2008. While there exist other algorithmic solutions for community detection, the Louvain algorithm has some distinct strengths and benefits, such as its speed, accuracy and suitability for large networks \citep{lancichinetti2012consensus}. Louvain works by firstly optimising modularity in smaller, local communities before combining local communities to larger ones. The algorithm runs until a stopping criteria is reached, typically when there is no further improvement in modularity. 

Although modularity optimisation using Louvain is applied extensively in various disciplines, it comes with a set of drawbacks that are particularly relevant for dealing with financial correlation matrices. One of these drawbacks is the well-known resolution limit problem \citep{fortunato2007resolution}. This problem describes the inability of modularity optimisation to detect communities that are smaller than a certain size. This is particularly problematic for financial correlation matrices, as we would expect to find communities of different, potentially small sizes.

An improvement to the Louvain algorithm, the so-called Leiden algorithm was proposed by \citep{traag2019louvain}. It not only overcomes the problem of arbitrarily badly connected communities, it is also capable of finding better solutions in a faster runtime. This algorithm implements the modularity quality function as defined in Function \eqref{eq_modularity} above. Leiden allows for an initial community membership of each node to a basic ground truth. Implementing an initial membership comes with benefits such as enhanced convergence speeds and increased robustness of the resulting community structures. By incorporating prior knowledge of node affiliations — such as sectoral data in financial networks — the Leiden algorithm can better align the detected communities with inherently meaningful groupings. This preliminary assignment of nodes to sectors helps to ground the algorithm's iterations, potentially reducing the number of steps needed to achieve a high-quality partitioning of the network. Leiden introduces three different run phases.

\begin{description}
   \item[Local moving of nodes] The algorithm begins with a local moving phase similar to the Louvain algorithm. It improves Louvain by applying a "fast local move" procedure. 
   \item[Refinement of the partition] Following creating the initial partition $P$, Leiden refines this partition by optimising each community in the partition as follows, creating a new partition $P_{refined}$.
   \item[Aggregation of network] In the final phase of the algorithm, Leiden aggregates the two partitions $P$ and $P_{refined}$.
\end{description}

These steps work in conjunction until the objective, an optimised modularity, is obtained. By following these three phases, Leiden guarantees well-connected communities as well as higher-quality partitions compared to Louvain.

Despite that Louvain has been used for various financial problems such as correlation clustering (see, e.g., \citep{macmahon2015community} and \citep{wang2017multiscale}), sentiment analysis in financial news networks \citep{wan2021sentiment} and systemic stability of markets during financial crisis \citep{heiberger2014stock}, Leiden has not yet had its breakthrough in the financial domain. Though based on the results of \citet{traag2019louvain}, we expect Leiden to outperform Louvain in our correlation clustering problem as well and implement it as part of our Dynamic Modularity-Spectral Algorithm. 

\section{Managing community sizes using spectral clustering}

Besides community detection algorithms, spectral clustering has been recently applied to financial clustering problems. 

While both modularity optimisation and spectral clustering are graph clustering techniques, they differ significantly in their methodology. As we have seen above, modularity optimisation aims to maximise (or minimise) a global quality function that describes the likelihood of a node being in a community \citep{newman2004finding}. This method has been shown to be effective in finding well-connected communities within graphs \citep{traag2019louvain}.

Spectral clustering on the other hand does not optimise a global criteria, but transforms the problem into a graph partitioning problem in a lower dimension \citep{shi2000normalized}. In other words, we can consider spectral clustering as a bottom up approach that considers the local structure of the graph by analysing pairwise similarities between nodes. Community detection on the other side has a global top down objective, which is a more holistic view of the overall graph structure.

\subsection{Spectral clustering similarity matrix}

To apply the spectral clustering algorithm, we have to transform the correlation based network into a similarity matrix. From this matrix, the Laplacian is calculated and the eigenvectors are extracted. Inspired by the unweighted pair group method with arithmetic mean (UPGMA) \citep{Sokal1958ASM} we calculate the similarity matrix as follows

\begin{equation}
	S_{i,j} = \frac{1}{N_{C_1} \times N_{C_2}} \sum_{a \in C_1} \sum_{b \in C_2} \rho(a,b)
\end{equation}

where $N_{C_1}$ and $N_{C_2}$ are the number of stocks in cluster $C_1$ and $C_2$ respectively and $\rho(a,b)$ is the correlation between stock $a$ and $b$ from the correlation matrix.

We opted for this method over the more standard Gaussian approach \citep{von2007tutorial} as we wanted to include the information from the community detection algorithm, namely the detection of smaller clusters. The Gaussian approach would have dropped this information and compared the individual stocks rather than the small clusters directly.

After receiving these new clusters, we compare them to the initial results from the Leiden algorithm by computing the differences 

\begin{equation}
	\Delta_{spectral} = \overline{\rho}_{intra_s}-\overline{\rho}_{inter_s}
\end{equation}

and 

\begin{equation}
	\Delta_{leiden} = \overline{\rho}_{intra_l}-\overline{\rho}_{inter_l}
\end{equation}

where $\overline{\rho}_{intra}$ is the average intra-cluster correlation and $\overline{\rho}_{inter}$ is the average inter-cluster correlation for all clusters obtained by the respective method Leiden or spectral. 

If $\Delta_{spectral} > \Delta_{leiden}$, we keep the spectral clustering result, otherwise we keep the Leiden result.

\subsection{Number of cluster constraints}

Unlike the community detection we implemented with the Leiden algorithm, spectral clustering requires a number of clusters a priori. This is due to the underlying clustering methodology, which in our case is based on standard k-means \citep{von2007tutorial}. While this is a limitation to the model we wanted to avoid by using community detection, it can also help to deal with certain edge cases. If we assume, that the Leiden algorithm assigns every stock to its own cluster, we ultimately remain at the starting point of our problem. Even if Leiden finds, e.g., 80 clusters, a few of these clusters being large and rest containing only 1 or 2 stocks, then we struggle to build standardised portfolios out of this. Therefore, reducing the number of clusters to a smaller number can help in the decision making process. However, we have to remain cautious in defining the number of clusters, as we do not want to lose the information from the community detection algorithm. Assuming that both spectral clustering and community detection are actually not capable of merging several nodes in clearly defined clusters, then these small clusters can potentially account for an increased diversification when selecting stocks for portfolio and we would want to select them first. Therefore we opt for a dynamic approach, in which the algorithm automatically tests for different numbers of clusters and selects the solution that results in a higher $\Delta$ as defined above. We arbitrarily limit the search space to a total number of clusters between 

\begin{equation}
	\textit{min\_clusters} = 5
\end{equation}
and
\begin{equation}
	max\_clusters = 
        \begin{cases}
            10, & \text{if } \mid{communities}\mid >10
            \\
            \mid{communities}\mid, & \text{otherwise }
        \end{cases}
\end{equation}

where $\mid{communities}\mid$ is the number of small communities found by the Leiden algorithm.

\section{Risk minimisation through stock selection}
As mentioned before, we assume that by applying graph cluster algorithms, we can detect a hidden market structure that offers investment opportunities for inherent diversification. We can assume that $\{C_1, C_2, \ldots, C_k\}$ are the clusters formed from the initial set of stocks $S$. Each cluster $C_i$ is evaluated for intra-cluster correlation $\rho_{\text{intra}}$ and inter-cluster correlation $\rho_{\text{inter}}$.

Mathematically, the risk $R(P)$ of the portfolio can be expressed as:
\begin{equation}
    R(P) = \sum_{i=1}^n w_i^2 \sigma_i^2 + \sum_{i=1}^n \sum_{j=1, j \neq i}^n w_i w_j \sigma_i \sigma_j \rho_{ij}
\end{equation}
where $w_i$ is the weight of stock $ i$ in the portfolio, and $\sigma_i$ is the standard deviation of stock $i$, representing the individual risk measured as volatility of an individual stock $i$. $\rho_{ij}$ is the correlation between stocks $i$ and $j$. The risk of the portfolio $R(P)$ consists of two components:
\begin{enumerate}
    \item The first term $\sum_{i=1}^n w_i^2 \sigma_i^2$ sums the weighted variances of all stocks in the portfolio $P$. It represents the contribution of the individual variances (risks) of the stocks to the overall portfolio risk.
    \item The second term $\sum_{i=1}^n \sum_{j=1, j \neq i}^n w_i w_j \sigma_i \sigma_j \rho_{ij}$ accounts for the weighted pairwise covariances between different stocks in the portfolio. The covariance $w_i w_j \sigma_i \sigma_j \rho_{ij}$ captures how the returns of stocks $i$ and $j$ co-move.
\end{enumerate}

By reducing the correlation term $\rho_{ij}$, we directly reduce the overall covariance of stocks in the portfolio, which reduces the portfolio risk $R(P)$.

\subsection{Selection of Stocks from Small and Large Clusters}
Based on our hypothesis four that stocks that are not perfectly attributable to clusters are better diversifiers, we have to select stocks in these peripheral areas of our graph first during the portfolio construction process. Additionally, we assume that stocks that have a lower intra-cluster correlation are also better diversifiers than stocks with the highest intra-cluster correlation. Therefore, we propose the following stock selection process:

\begin{enumerate}
    \item Select all stocks from small clusters where $n$ defines the size of small clusters:
    \begin{equation}
        S_{\text{small}} = \bigcup_{C_i \leq n} C_i
    \end{equation}
    \item Select stocks from larger clusters proportionally to their size:
    \begin{equation}
        S_{\text{large}} = S \setminus S_{\text{small}}
    \end{equation}

    \begin{equation}
        m = \sum_{C_j > n} C_j
    \end{equation}

    \begin{equation}
        S_{\text{large}}^{\text{sorted}} = \bigcup_{C_j > n} \text{sort}(C_j, \rho_{\text{intra}})
    \end{equation}

    \begin{equation}
        S_{\text{large}}^{\text{selected}} = \bigcup_{C_j > n} \left( \text{select } \left\lfloor \frac{C_j}{m} \times (k - S_{\text{small}}) \right\rfloor \text{ stocks with lowest } \rho_{\text{intra}} \text{ from } C_j \right)
    \end{equation}

    where $k$ is the number of stocks included in the portfolio.

    \item The final diversified set of $k$ stocks:
    \begin{equation}
        S_k = S_{\text{small}} \cup S_{\text{large}}^{\text{selected}}
    \end{equation}

\end{enumerate}

Here, the function $\text{sort}(C_j, \rho_{\text{intra}})$ sorts the stocks within cluster $C_j$ in ascending order based on their intra-cluster correlation $\rho_{\text{intra}}$. Mathematically, this can be defined as:

\begin{equation}
    \text{sort}(C_j, \rho_{\text{intra}}) = \left( s_{C_j,1}, s_{C_j,2}, \ldots, s_{C_j,|C_j|} \right)
\end{equation}

where $s_{C_j,i}$  and $s_{C_j,i+1}$ are stocks in cluster $C_j$ such that $\rho_{\text{intra}}(s_{C_j,i}) \leq \rho_{\text{intra}}(s_{C_j,i+1})$ and $|C_j|$ defines the cardinality of $C_j$. This sorting ensures that stocks with the lowest intra-cluster correlation are considered first for selection.

To test our assumption that these stocks are better diversifiers and yield stronger portfolios, we additionally build portfolios where we select the stocks with the highest intra-cluster correlation.

\section{Application of DynMSA to US Equities}
\subsection{Data}
The data we use to compute the correlation matrix are daily closing prices for S\&P 500. For the initial correlation clustering based on daily returns, we consider the full data set of current constituents that were members of these indices since the start of our data set in July 2017. This leaves us with 462 stocks for the S\&P 500. We compute daily returns by 

\begin{equation}
r_{t} = \frac{P_{t} - P_{t-1}}{P_{t-1}}
\end{equation}

where $P$ is the daily closing price.

Additionally we compute an equally weighted performance benchmark which is also rebalanced monthly. 

\subsection{Similarity of clusters}
To analyse our clusters beyond our objective and the subsequent portfolio allocation, we are also interested in some statistical measures of our clusters. Particularly, we are interested in finding common market structures and regime changes. If we assume that the clusters derived from our DynMSA represent the market structure and the general market structure being stable over time, then the algorithm should cluster the same stocks together in each period. This should change particularly when there is a regime shift or exogenous shocks like the financial crisis or COVID-19 pandemic. The Adjusted Rand Index (ARI) is a method to measure the similarity of two clusterings, even when the number of clusters and cluster sizes changes over time \citep{hubert1985comparing}. The ARI is particularly well-suited for evaluating the consistency and changes in inter-temporal market structures due to its normalisation for chance grouping.

The ARI is an adjusted version of the original Rand Index \citep{rand1971objective}. It can be expressed as follows:

\begin{equation}
    \label{eq_ari}
        ARI = \frac{RI-Expected_{RI}}{Max_{RI}-Expected_{RI}}
\end{equation}

where $RI$ is the Rand Index, $Expected_{RI}$ is the expexted RI under random chance and $Max_{RI}$ is the maximum possible value of Rand Index. The Rand Index is defined as 

\begin{equation}
    \label{eq_ri}
        ARI = \frac{a+d}{a+b+c+d}
\end{equation}

where $a$ is the number of pairs of elements in the same cluster in both clusterings, $b$ is the number of pairs of elements that are in the same cluster in the first clustering but in different clusters in the second, $c$ is the number of pairs of elements that are in different clusters in the first clustering but in the same cluster in the second clustering and lastly $d$ are pairs of elements that are in different clusters in both clusterings.

The ARI typically has values between $0$ for completely random clusters and $1$ if the clusters are perfectly identical \citep{hubert1985comparing}. We use the ARI to compare the stock clusters over time. If the market structure remains similar, we expect higher values for ARI while we would expect drops in the ARI whenever we encounter regime changes or exogenous shocks.

\subsection{Community analysis}
In this section, we analyse the clusters obtained by our ensemble approach. We compare our results with a predefined baseline, in our case clusters based on the GICS sector classification of our primary index. Additionally, we evaluate the performance of our clustering approach against a market-based benchmark, which includes all stocks in our investment universe equally weighted.

\subsubsection{Finding the optimal correlation look-back time frame}
A critical aspect of our analysis involves examining the influence of different time frames for calculating the correlation look-backs. The selection of an appropriate time frame can heavily influence the accuracy and reliability of the clustering process. Selecting an appropriate time frame is not trivial as there is no clear one-fit solution proposed in the academic literature. However, \citet{marti2016clustering} suggest a time frame between 250 and 500 trading days based on the number of stocks in the data set and clustering methodology that they tested. For the upper bound, we orientate ourselves on these time frames. We include shorter time frames as well as they could adjust the clusters more dynamically and direct once a market shift occurs. 

By analysing different time frames, we aim to observe how the intra-cluster and inter-cluster correlation behave under varying scenarios. Identifying a look-back period that maximises the intra-cluster correlation while minimising the inter-cluster correlation could be considered optimal. We compare these values with the sector-based baseline to see if our algorithm can improve a naive clustering method that is yet widely used in the industry.

\begin{figure}
    \centering
    \begin{subfigure}[b]{0.45\textwidth}
        \centering
        \includegraphics[width=\textwidth]{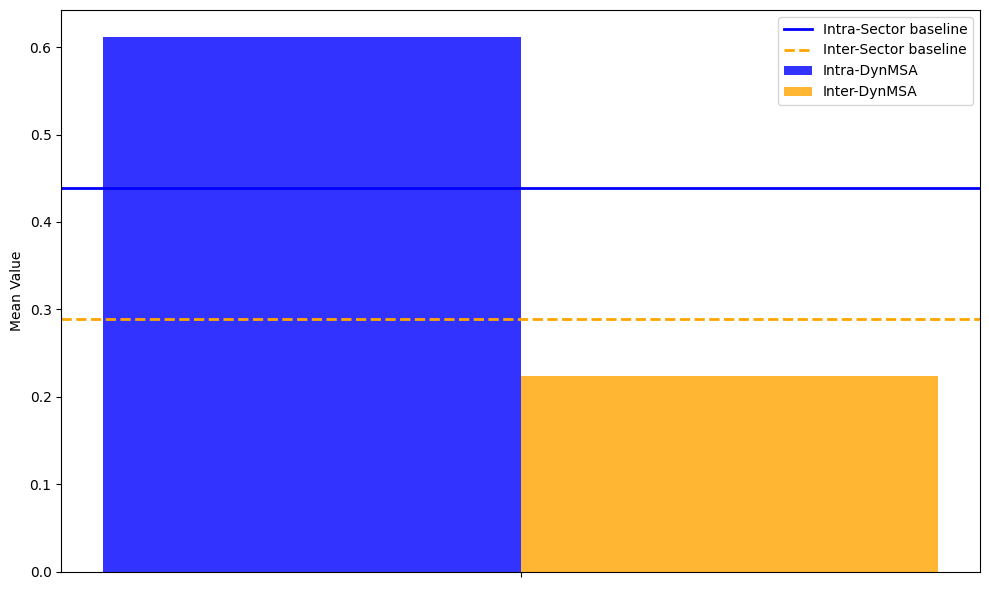}
        \caption{3-months look-back period}
        \label{fig:plot1}
    \end{subfigure}
    \hfill
    \begin{subfigure}[b]{0.45\textwidth}
        \centering
        \includegraphics[width=\textwidth]{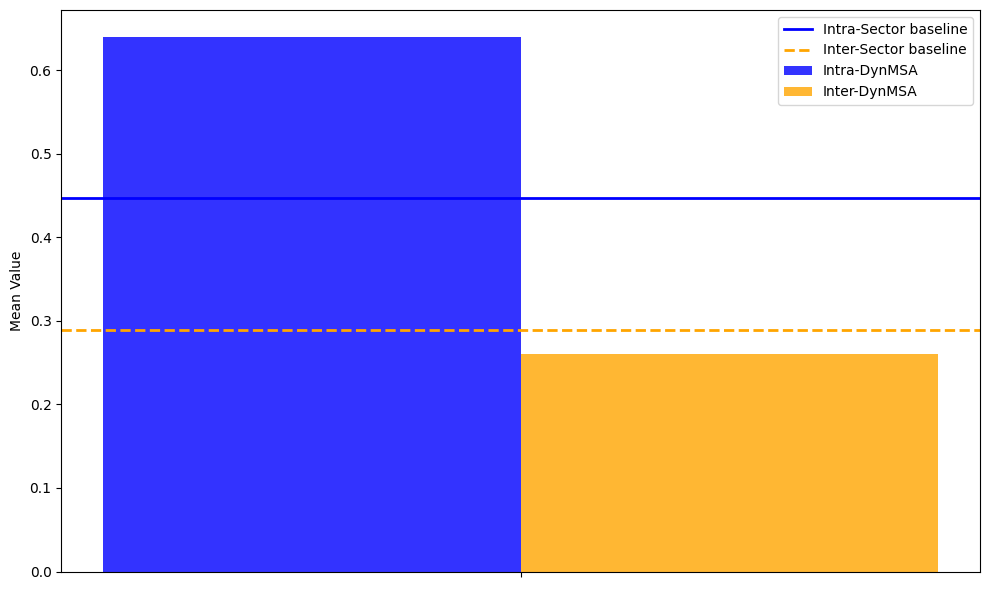}
        \caption{6-months look-back period}
        \label{fig:plot2}
    \end{subfigure}
    
    \vspace{0.5cm} 
    
    \begin{subfigure}[b]{0.45\textwidth}
        \centering
        \includegraphics[width=\textwidth]{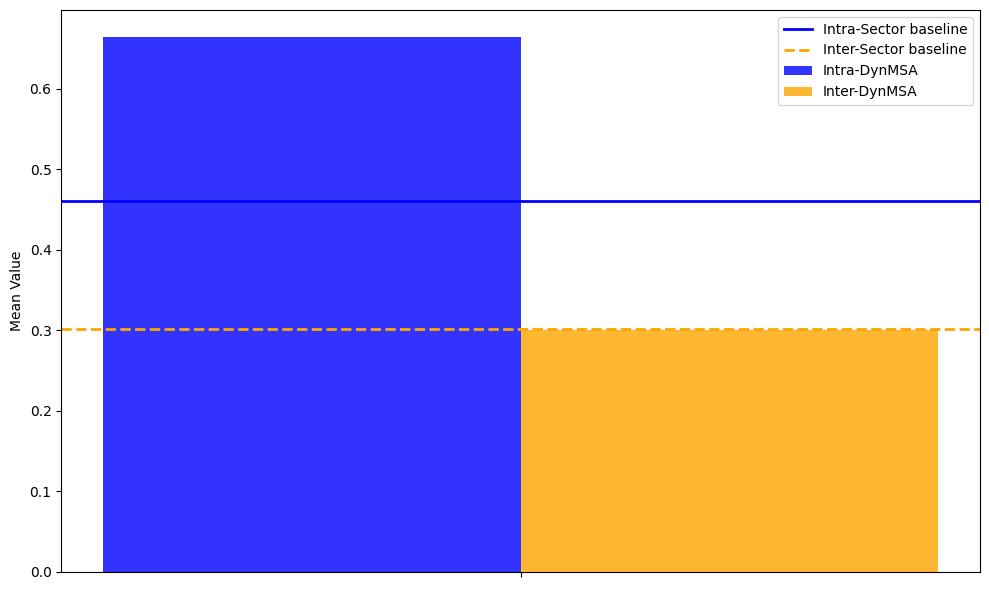}
        \caption{12-months look-back period}
        \label{fig:plot3}
    \end{subfigure}
    \hfill
    \begin{subfigure}[b]{0.45\textwidth}
        \centering
        \includegraphics[width=\textwidth]{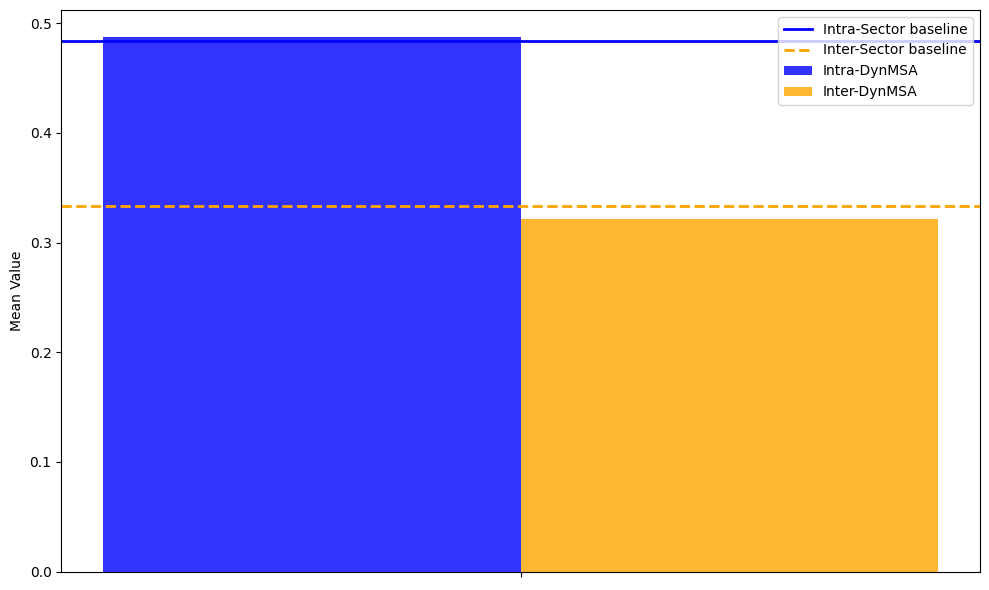}
        \caption{24-months look-back period}
        \label{fig:plot4}
    \end{subfigure}
    \caption{Comparison of average intra- and inter-cluster correlations over different look-back periods against the sector-based benchmark cluster.}
    \label{fig:comparison_lookback}
\end{figure}

For the 3-months look-back period, the intra-cluster correlation for our clustering method is significantly higher, with values around 0.6, compared to the intra-sector baseline, which is closer to 0.4. This difference indicates that the clusters effectively capture short-term relationships among stocks. The inter-cluster correlation for the clustering method is approximately 0.2, which is lower than the inter-sector baseline value of around 0.3. This lower inter-cluster correlation demonstrates that the clusters are well-separated from each other, making the clustering method highly effective for differentiating between various market relationships in the short term.

The results for the 6-months look-back period show a similar trend. The intra-cluster correlation remains high, around 0.6, outperforming the intra-sector baseline. This indicates that the method continues to capture meaningful short-term relationships over a slightly longer horizon. The inter-cluster correlation for the 6-months period is around 0.2, remaining lower than the inter-sector baseline of about 0.3. As before, our cluster algorithm outperforms the sector-based benchmark significantly.

When extending the look-back period to 12 months, the intra-cluster correlation remains high, though slightly lower at around 0.55, compared to the shorter periods. The intra-sector baseline of about 0.4 means that the intra-cluster correlation still significantly exceeds the baseline. The inter-cluster correlation is around 0.25, slightly higher than in the shorter periods but still below the inter-sector baseline of 0.3. While still outperforming the baseline, at least the inter-cluster correlation is now more aligned. This could be interpreted in a way that over 12 months, our algorithm cannot find well-distinguished clusters that are significantly different in terms of their correlations than others.

For the 24-months look-back period, the intra-cluster correlation decreases to around 0.5 but remains above the intra-sector baseline of approximately 0.4. This further decrease indicates that the method captures long-term relationships less effectively but maintains a good level of intra-cluster cohesion. The inter-cluster correlation for the 24-months period is around 0.3, higher than in the shorter periods but still below the inter-sector baseline. In combination with the 12 months results this suggests that while the clusters are distinct, the differentiation is less pronounced over the longer period. The 24-months period shows a reduced ability to maintain distinct clusters, reflecting the challenge of capturing long-term market relationships while ensuring cluster distinctiveness.

The analysis shows that shorter look-back periods, 3 and 6 months, are more effective in capturing and maintaining distinct market relationships, as evidenced by higher intra-cluster correlations and lower inter-cluster correlations. The medium-term look-back period of 12 months maintains high intra-cluster correlations and low inter-cluster correlations. The long-term period, while still effective with intra-cluster correlations around 0.5 and inter-cluster correlations around 0.3, shows a reduced ability to differentiate clusters distinctly. This analysis indicates that the shorter the look-back, the better the results. We will consider this finding in our further analysis and see, if it holds up. 

In this section we also tested and answered part of our first hypothesis, namely if DynMSA can lead to clusters with a significantly higher intra-cluster correlation and lower inter-cluster correlation than sector-based clusters. After our first subroutine, the modularity optimisation, this clearly is the case. The next section will test if hypothesis two stands, namely that DynMSA is better than modularity optimisation on its own. If this is the case, we can also conclude that our first hypothesis stands.

\subsubsection*{Improvements based on spectral clustering}
This section evaluates if and how the addition of spectral clustering to community detection in our DynMSA methodology improves the clustering results. The effectiveness of our approach is assessed through various metrics such as changes in intra-cluster correlation, inter-cluster correlation, modularity and our objective function \ref{eq_opt_objective}. The analysis is based on the average number of clusters before and after applying spectral clustering, as well as the numerical differences, percentage changes, standard deviations and statistical significance of improvements.

\begin{table}
\centering
\begin{adjustbox}{max width=\textwidth}
\begin{tabular}{ccc|ccc}
\toprule
 \multicolumn{3}{c|}{3 Months} & \multicolumn{3}{c}{6 Months} \\
 \midrule
 \textbf{initial better} & \textbf{equal} & \textbf{updated better} & \textbf{initial better} & \textbf{equal} & \textbf{updated better} \\
 \midrule
  0 & 63 & 18 & 0 & 69 & 9 \\
\midrule
 \multicolumn{3}{c|}{12 Months} & \multicolumn{3}{c}{24 Months} \\
 \midrule
 \textbf{initial better} & \textbf{equal} & \textbf{updated better} & \textbf{initial better} & \textbf{equal} & \textbf{updated better} \\
 \midrule
  0 & 44 & 28 & 0 & 60 & 0 \\
\bottomrule
\end{tabular}
\end{adjustbox}
\caption{Comparison of initial, equal, and updated clusters for different look-back periods}
\label{tab:comparison_all_periods}
\end{table}

As we can see in Table \ref{tab:comparison_all_periods}, except for 24 months look-back, spectral clustering improved the objective in each period despite stable average numbers of clusters. For the 3-months look-back period, spectral clustering improves the objective function in 18 cases, while the results remain equal in 63 cases. There are no instances where the initial clustering performs better than the updated clustering. For the 6-months look-back period, spectral clustering improves the objective function in 9 cases, with the results being equal in 69 cases. Similar to the 3-months period, there are no cases where the initial clustering is better than the updated clustering. This consistency demonstrates that spectral clustering continues to provide benefits by improving the clustering objective, even though with fewer improvements compared to the 3-months period.

In the  12-months period, spectral clustering improves the objective function in 28 cases, much more than the two shorter periods. In 44 cases the results are equal and analogue to the previous results, there are no cases where the initial clustering was better. Contrary to these results, there are no improvements or declines for the 24-months period. This could simply mean that for a longer look-back, the community detection algorithm already found approximately optimal clusters, leaving no room for improvements that spectral clustering could harness. 

Further analysis of the improvements can be seen in the Tables \ref{tab:summary-differences} and \ref{tab:t-test-summary}. 

\begin{table}[H]
\centering
\begin{adjustbox}{max width=\textwidth}
\begin{tabular}{c}
    \begin{subtable}[t]{\linewidth}
        \centering
        \begin{adjustbox}{max width=\textwidth}
        \begin{tabular}{lrrrr}
        \toprule
        Period & Mean Intra Corr Diff & Mean Inter Corr Diff & Mean Modularity Diff & Mean Objective Diff \\
        \midrule
        3 Months & 0.0040 & -0.0093 & -0.0000 & 0.0134 \\
        6 Months & -0.0003 & -0.0079 & -0.0000 & 0.0076 \\
        12 Months & 0.0556 & 0.0102 & -0.0000 & 0.0454 \\
        24 Months & 0.0000 & 0.0000 & 0.0000 & 0.0000 \\
        \bottomrule
        \end{tabular}
        \end{adjustbox}
        \caption{Mean Differences}
        \label{tab:mean-differences}
    \end{subtable} \\
    \vspace{0.2cm} 
    \begin{subtable}[t]{\linewidth}
        \centering
        \begin{adjustbox}{max width=\textwidth}
        \begin{tabular}{lrrrr}
        \toprule
        Period & Intra Corr Pct Change & Inter Corr Pct Change & Modularity Pct Change & Objective Pct Change \\
        \midrule
        3 Months & 0.7038 & -3.1814 & -0.0002 & 4.7173 \\
        6 Months & -0.0789 & -2.3008 & -0.0001 & 3.6738 \\
        12 Months & 9.5249 & 4.0391 & -0.0018 & 13.5253 \\
        24 Months & 0.0000 & 0.0000 & 0.0000 & 0.0000 \\
        \bottomrule
        \end{tabular}
        \end{adjustbox}
        \caption{Percentage Changes}
        \label{tab:percentage-changes}
    \end{subtable} \\
    \vspace{0.2cm} 
    \begin{subtable}[t]{\linewidth}
        \centering
        \begin{adjustbox}{max width=\textwidth}
        \begin{tabular}{lrrrr}
        \toprule
        Period & Intra Corr Std Dev & Inter Corr Std Dev & Modularity Std Dev & Objective Std Dev \\
        \midrule
        3 Months & 0.0278 & 0.0236 & 0.0000 & 0.0364 \\
        6 Months & 0.0082 & 0.0250 & 0.0000 & 0.0241 \\
        12 Months & 0.0804 & 0.0199 & 0.0000 & 0.0703 \\
        24 Months & 0.0000 & 0.0000 & 0.0000 & 0.0000 \\
        \bottomrule
        \end{tabular}
        \end{adjustbox}
        \caption{Standard Deviations}
        \label{tab:standard-deviations}
    \end{subtable}
\end{tabular}
\end{adjustbox}
\caption{Summary of differences, percentage changes, and standard deviations}
\label{tab:summary-differences}
\end{table}

Table \ref{tab:summary-differences} highlights the absolute differences before and after running the spectral clustering, as well as the percentage change in standard deviation. In Table \ref{tab:summary-differences}(a), for the 3-months period, there is a slight increase in the mean intra-cluster correlation and a decrease in the mean inter-cluster correlation, indicating improved cohesion within clusters and better separation between clusters. The 6-months period shows similar trends, with a minor decrease in the mean intra-cluster correlation and a decrease in the mean inter-cluster correlation, alongside an improvement in the objective function. The 12-months period shows significant improvements, with a substantial increase in the mean intra-cluster correlation and a slight increase in the mean inter-cluster correlation, resulting in a notable improvement in the objective function. The 24-months period shows no changes, with all metrics remaining at zero as expected based on the previous results in Table \ref{tab:comparison_all_periods}.

Table \ref{tab:summary-differences}(b) shows the percentage changes in the same metrics. The 3-months period shows a slight increase in intra-cluster correlation and a decrease in inter-cluster correlation, with the objective function improving. The 6-months period shows minor changes with improvements in the objective function. The 12-months period shows significant changes, with an increase in both intra-cluster and inter-cluster correlation, leading to a notable improvement in the objective function. These results highlight improvements in the objective function across all periods, with the 12-months look-back showing the most significant improvement, driven by a large increase in intra-cluster correlation offsetting a slight increase in inter-cluster correlation.

In terms of modularity, the objective for the initial community detection shows a decline over all look-back periods where spectral clustering adjusted the clusters. This makes intuitive sense as the cluster composition changes, even though the initial clustering was already strong, indicating that the modularity function inherently aligns with the objective function defined for spectral clustering.

\begin{table}
\centering
\begin{adjustbox}{max width=\textwidth}
\begin{tabular}{c}
    \begin{subtable}[t]{\linewidth}
        \centering
        \begin{adjustbox}{max width=\textwidth}
        \begin{tabular}{lrrrr}
        \toprule
        Period & Intra Corr t-test Stat & Intra Corr p-value & Inter Corr t-test Stat & Inter Corr p-value \\
        \midrule
        3 Months & 1.3047 & 0.1957 & -3.5618 & 0.0006 \\
        6 Months & -0.3184 & 0.7511 & -2.7899 & 0.0066 \\
        12 Months & 5.8675 & 0.0000 & 4.3500 & 0.0000 \\
        24 Months & NaN & NaN & NaN & NaN \\
        \bottomrule
        \end{tabular}
        \end{adjustbox}
        \caption{t-test statistics and p-values for correlations}
        \label{tab:correlation-t-test-pvalues}
    \end{subtable} \\
    \vspace{0.2cm} 
    \begin{subtable}[t]{\linewidth}
        \centering
        \begin{adjustbox}{max width=\textwidth}
        \begin{tabular}{lrrrr}
        \toprule
        Period & Modularity t-test Stat & Modularity p-value & Objective t-test Stat & Objective p-value \\
        \midrule
        3 Months & -1.4509 & 0.1507 & 3.3069 & 0.0014 \\
        6 Months & -1.0603 & 0.2923 & 2.7901 & 0.0066 \\
        12 Months & -2.5257 & 0.0138 & 5.4792 & 0.0000 \\
        24 Months & NaN & NaN & NaN & NaN \\
        \bottomrule
        \end{tabular}
        \end{adjustbox}
        \caption{t-test statistics and p-values for modularity and objective}
        \label{tab:modularity-objective-t-test-pvalues}
    \end{subtable}
\end{tabular}
\end{adjustbox}
\caption{Summary of t-test statistics and p-values}
\label{tab:t-test-summary}
\end{table}

Tables \ref{tab:t-test-summary}(a) and \ref{tab:t-test-summary}(b) provide the t-test statistics and p-values for the differences in metrics, measuring the statistical significance of changes caused by spectral clustering. For the 3-months period, the reduction in inter-cluster correlation is highly significant, indicating a strong improvement in cluster separation. The objective function also shows significant improvement, while changes in intra-cluster correlation and modularity are not significant, suggesting that the structural integrity of the clusters remains stable.

In the 6-months look-back period, the reduction in inter-cluster correlation and improvement in the objective function remain highly significant. Changes in intra-cluster correlation and modularity are not significant, indicating stable modularity and intra-cluster relationships.

The 12-months period shows highly significant improvements across all metrics. The increase in intra-cluster correlation indicates a strong improvement in cluster cohesion. The reduction in inter-cluster correlation and improvement in the objective function are also highly significant. Changes in modularity are slightly significant, suggesting a minor decline, but the structural integrity of the clusters is maintained.

Lastly, the 24-months period shows no change, as expected, with clusters remaining the same after spectral clustering as those obtained by community detection.

Overall, spectral clustering enhances clustering results, particularly over shorter and medium-term periods. For the 3-months and 6-months periods, it significantly improves inter-cluster separation and the objective function without significantly affecting modularity. The 12-months period demonstrates highly significant improvements across all metrics, highlighting the effectiveness of spectral clustering in this medium-term period. The stability of modularity across all periods indicates that spectral clustering enhances clustering quality while preserving the structural integrity identified by the initial community detection algorithm.

With the results of this section, we can conclude that hypothesis one stands. DynMSA can find clusters that have a higher intra- and a lower inter-cluster correlation compared to sector-based clusters. Furthermore, our results show that the combined approach is equal to or better than modularity optimisation on its own, confirming our second hypothesis.

\subsection{Stability of clusters over time}
Here we compute the Adjusted Rand Index as described in Equation \ref{eq_ari} for the stocks. We plot all four correlation look-backs and compare the clusters on a rolling monthly basis. We also compute the z-scores to detect potential outliers. Additionally, we compare the total number of clusters for each look-back period over time. This section therefore checks our third hypothesis, namely that DynMSA detects stable clusters over time while also detecting regime changes.

\begin{figure}
	\includegraphics[width = \textwidth]{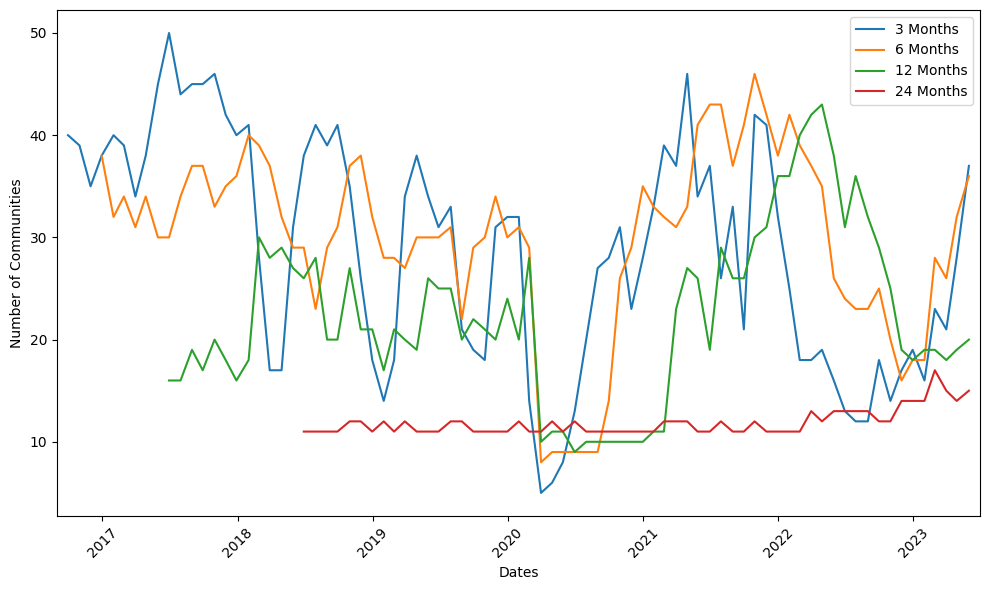}
	\caption{Number of clusters over time}
	\label{fig:comm_over_time_full}
\end{figure}

Starting with the number of clusters over time in Figure \ref{fig:comm_over_time_full}, we can see a drop in communities during March 2020 and following, clearly coinciding with the COVID-19 crisis. In terms of overall numbers of communities, the average number we computed above indicated that the two shorter periods, 3 and 6 months, have the highest number of clusters. We can visually see this in this plot that this holds truth for most months in our data set. However, during March 2020, the number of clusters dropped significantly below 10. This highlights the influence of COVID-19 as an exogenous shock on the market which led to a decline in most stock prices across all sectors. Contrary to this, the 24 months time series shows hardly any fluctuation over the whole data set. The number of clusters remains stable just over 10, slightly trending higher towards the end of the data set in the second half of 2022 and the first half of 2023. 

\begin{figure}
    \centering
    \begin{subfigure}[b]{0.45\textwidth}
        \centering
        \includegraphics[width=\textwidth]{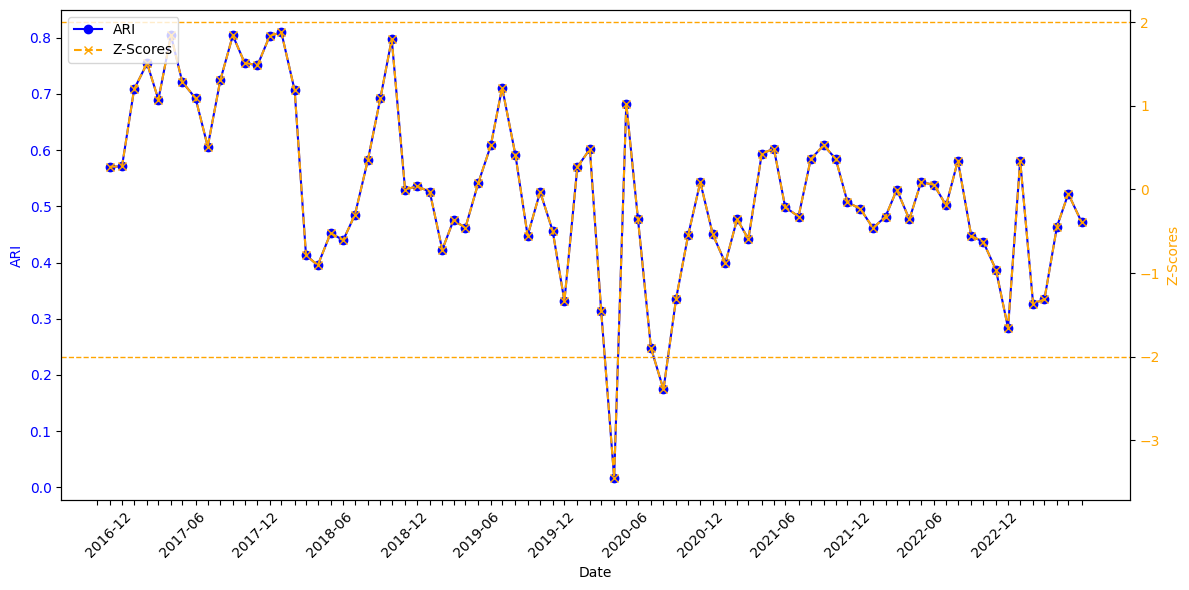}
        \caption{3-months look-back period}
        \label{fig:ari_zscore_3m}
    \end{subfigure}
    \hfill
    \begin{subfigure}[b]{0.45\textwidth}
        \centering
        \includegraphics[width=\textwidth]{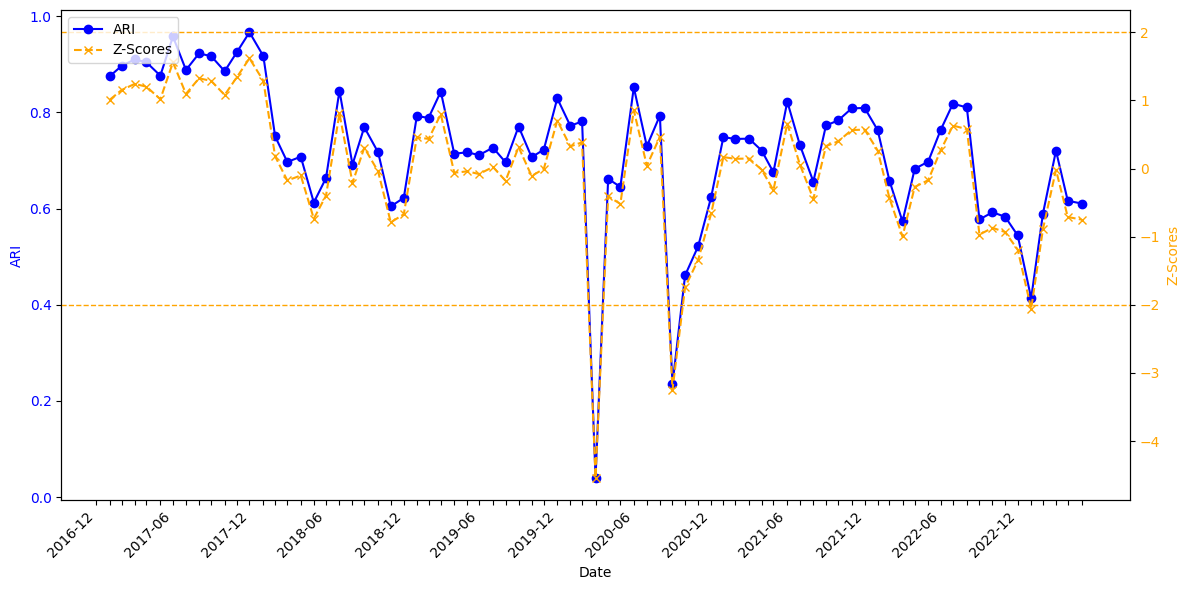}
        \caption{6-months look-back period}
        \label{fig:ari_zscore_6m}
    \end{subfigure}
    
    \vspace{0.5cm} 
    
    \begin{subfigure}[b]{0.45\textwidth}
        \centering
        \includegraphics[width=\textwidth]{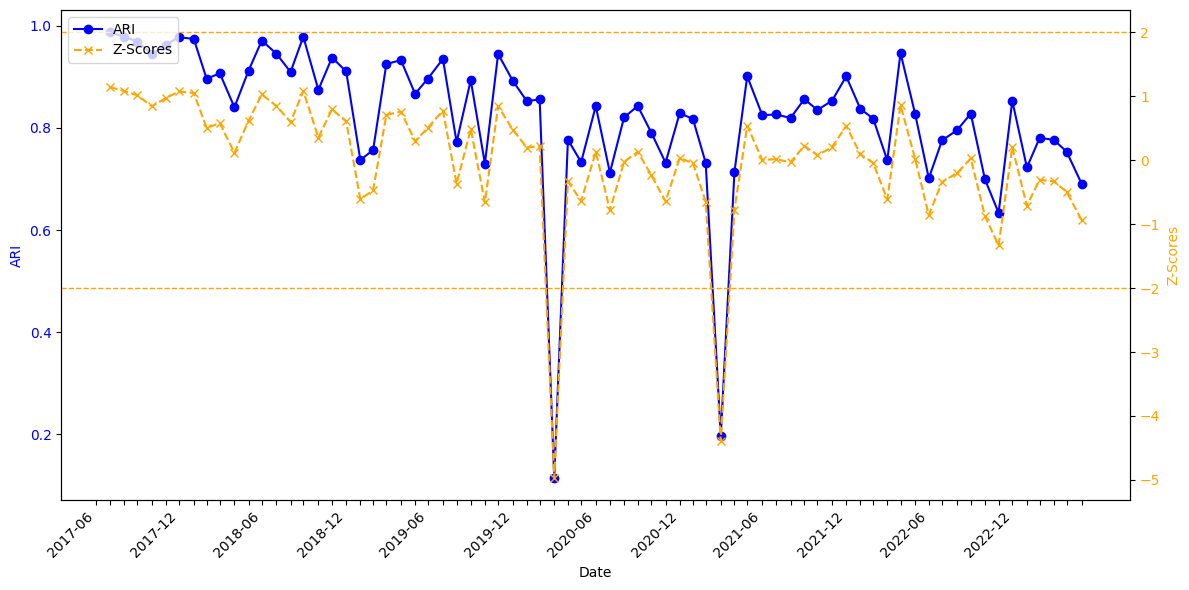}
        \caption{12-months look-back period}
        \label{fig:ari_zscore_12m}
    \end{subfigure}
    \hfill
    \begin{subfigure}[b]{0.45\textwidth}
        \centering
        \includegraphics[width=\textwidth]{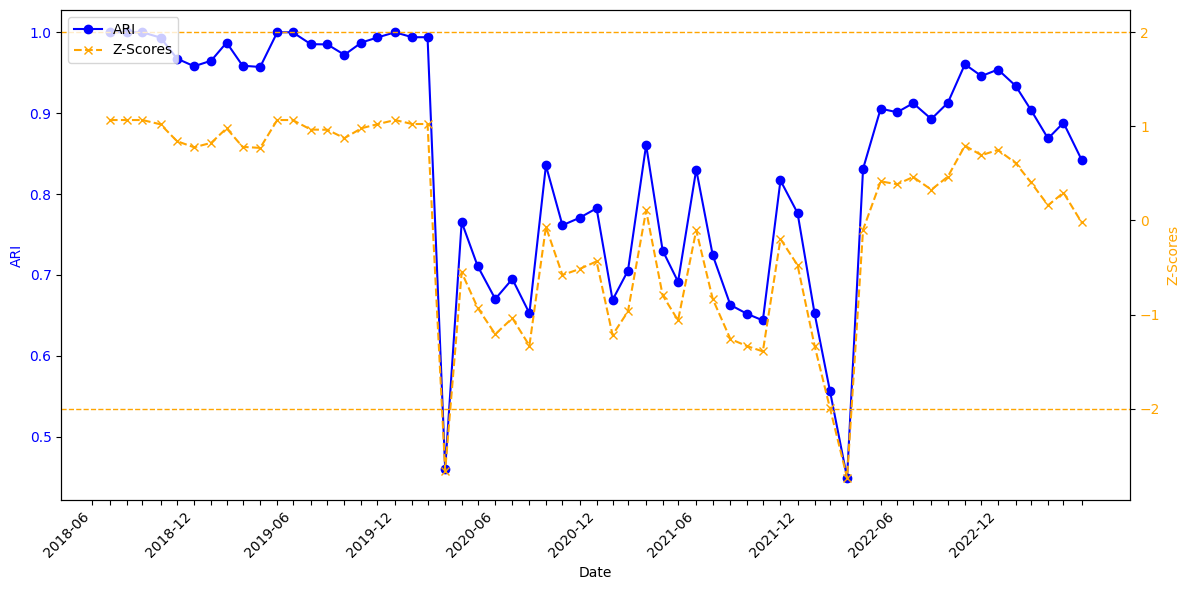}
        \caption{24-months look-back periods}
        \label{fig:ari_zscore_24m}
    \end{subfigure}
    \caption{ARI and z-score over time for different correlation look-back periods.}
    \label{fig:ari_zscore}
\end{figure}

Considering the impact of COVID-19 on stocks, we can note that the number of clusters for 3, 6 and 12 months seem to increase significantly again once the data for March 2020 leaves the look-back period. In Figure \ref{fig:ari_zscore} we can see the same phenomenon when computing the ARI score and the relevant Z-values.

For the 3-months look-back period, the ARI values exhibit substantial variability, frequently oscillating with notable fluctuations. This high degree of fluctuation indicates that the clustering method is highly sensitive to short-term market conditions, capturing rapid shifts in market structure. The z-scores corresponding to this period highlight significant deviations from the mean ARI, particularly around early 2020, aligning with the start of the COVID-19 pandemic. This period of market turbulence is marked by a pronounced drop in the ARI, reflecting the substantial impact of the pandemic on market structure.

The 6-months look-back period shows a slightly moderated variability in ARI values compared to the 3-months period. While still sensitive to market changes, the ARI fluctuations are less extreme, indicating a balance between short-term sensitivity and robustness. The z-scores for this period also reflect significant deviations but with less intensity than the 3-months period, suggesting that the 6-months look-back period captures market dynamics with a more moderate response.

The 12-months look-back period provides a much more stable and higher ARI than the previous two periods. While we can still see some market sensitivity from month to month, the fluctuations are less drastic than before. The ARI values point towards a stronger resilience to short-term noise and better capture medium-term structural changes. March 2020 remains as a significant, exogenous market event and after 12 months, we can see the COVID-19 observations leaving the data set. These larger fluctuations highlight a similarity in clusters before, during and after the influence of the COVID-19 shock being present in the data. Overall however, the z-scores show fewer extreme values, indicating a more stable clustering result over this medium-term period.

Lastly, the 24-months look-back exhibits the highest ARI values and stability outside the influence of the COVID-19 shock. Particularly at the start of the period we can see ARI values near 1, which means that the cluster algorithm returned nearly the same clusters every month. Even during COVID-19, the ARI scores remain higher and the z-scores lower than in the periods before, indicating the capabilities of capturing more long-term market trends. From an industrial point of view this could be interesting to implement in larger portfolios as more stability hints towards less turnover and rebalancing costs.

We can note that overall, shorter look-back periods, such as 3 and 6 months, are highly responsive to rapid market changes and capture short-term dynamics effectively. However, they also exhibit higher variability and sensitivity to market events which was particularly highlighted by the sharp drop in number of clusters caused by the influence of COVID-19 in March 2020. In contrast, longer look-back periods, such as 12 and 24 months, offer greater stability and are less affected by short-term noise, making them more suitable for capturing long-term structural trends in the market. This suggests that when implementing our algorithm, selecting the appropriate look-back period based on investment horizons and objectives can significantly enhance portfolio management practices, offering a sophisticated tool for strategic asset allocation and risk management.

\subsection{Stock cluster frequency and sectoral analysis}
To better understand the clusters derived from our methodology in terms of their constituents and their stability over time, in this section we analyse how often certain stocks are clustered together as well as the sector distributions per cluster over time. Hypothesis four asks exactly this question. Can DynMSA find a hidden market structure based on correlations that is different from the standard sector classification.

\begin{table}[H]
	\centering
		\begin{adjustbox}{max width=\textwidth}
			\begin{tabular}{lrll}
				\toprule
				\midrule
				0 & 10 & Energy & blue \\
				1 & 15 & Materials & orange \\
				2 & 20 & Industrials & green \\
				3 & 25 & Consumer Discretionary & red \\
				4 & 30 & Consumer Staples & purple \\
				5 & 35 & Health Care & brown \\
				6 & 40 & Financials & pink \\
				7 & 45 & Information Technology & grey\\
				8 & 50 & Communication Services & cyan \\
				9 & 55 & Utilities & yellow \\
				10 & 60 & Real Estate & lime \\
				\bottomrule
				\end{tabular}
		\end{adjustbox}
	\caption{Sector mapping}
	\label{tab:sector_mapping}
	\end{table}

\begin{figure}
    \centering
    \begin{subfigure}[b]{0.45\textwidth}
        \centering
        \includegraphics[width=\textwidth]{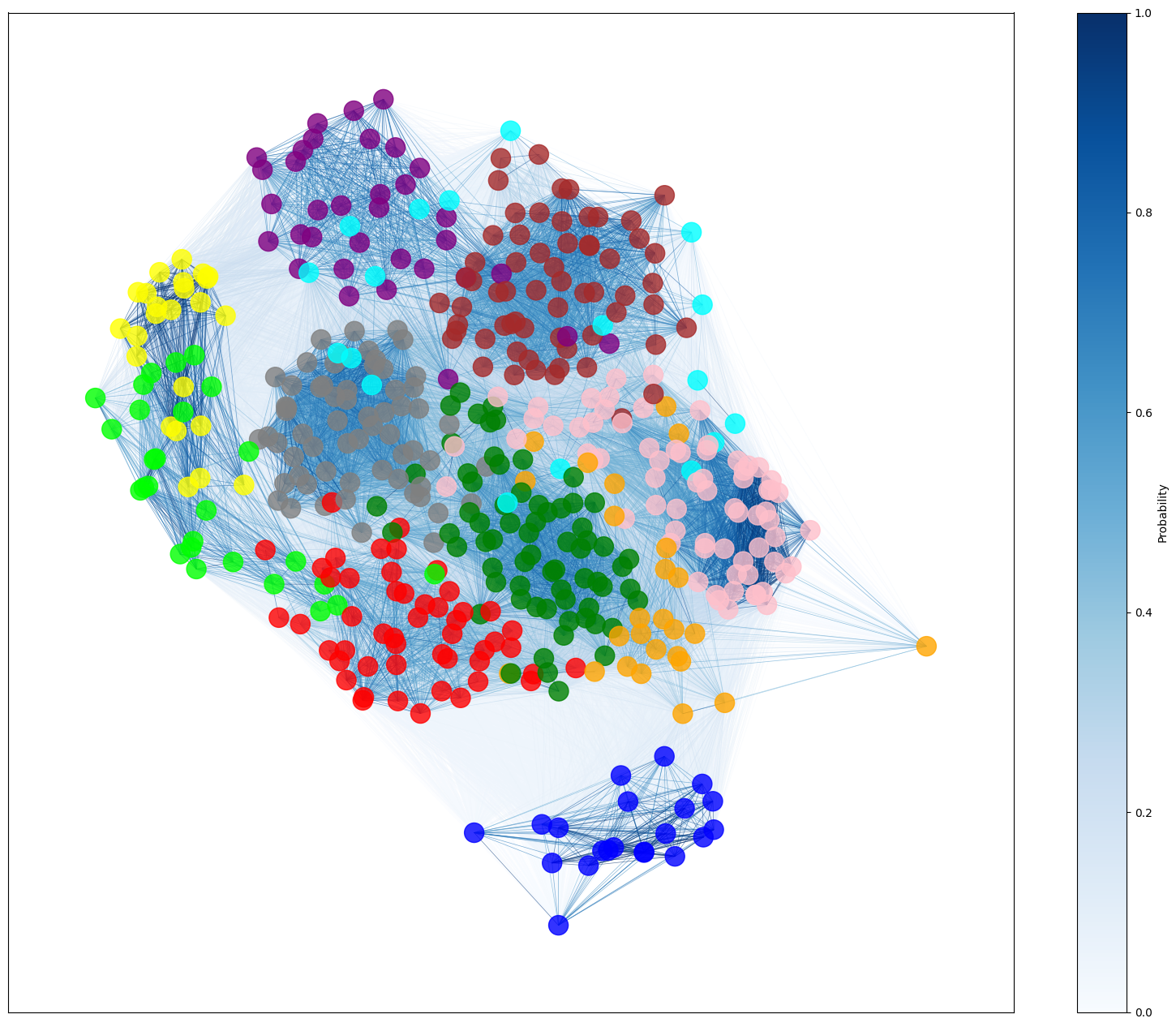}
        \caption{3-months look-back period}
        \label{fig:cluster_prob_3m}
    \end{subfigure}
    \hfill
    \begin{subfigure}[b]{0.45\textwidth}
        \centering
        \includegraphics[width=\textwidth]{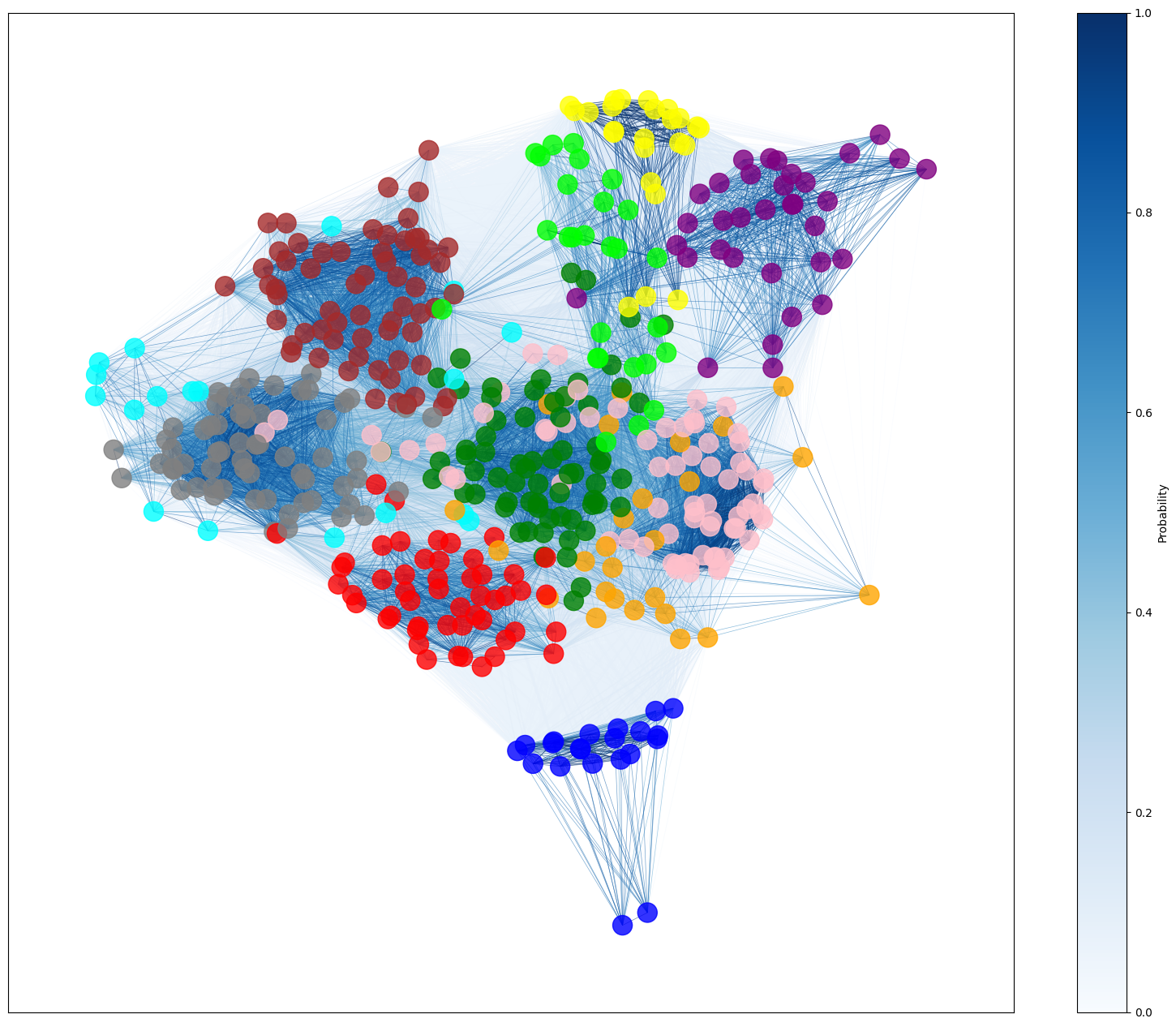}
        \caption{6-months look-back period}
        \label{fig:cluster_prob_6m}
    \end{subfigure}
    
    \vspace{0.5cm} 
    
    \begin{subfigure}[b]{0.45\textwidth}
        \centering
        \includegraphics[width=\textwidth]{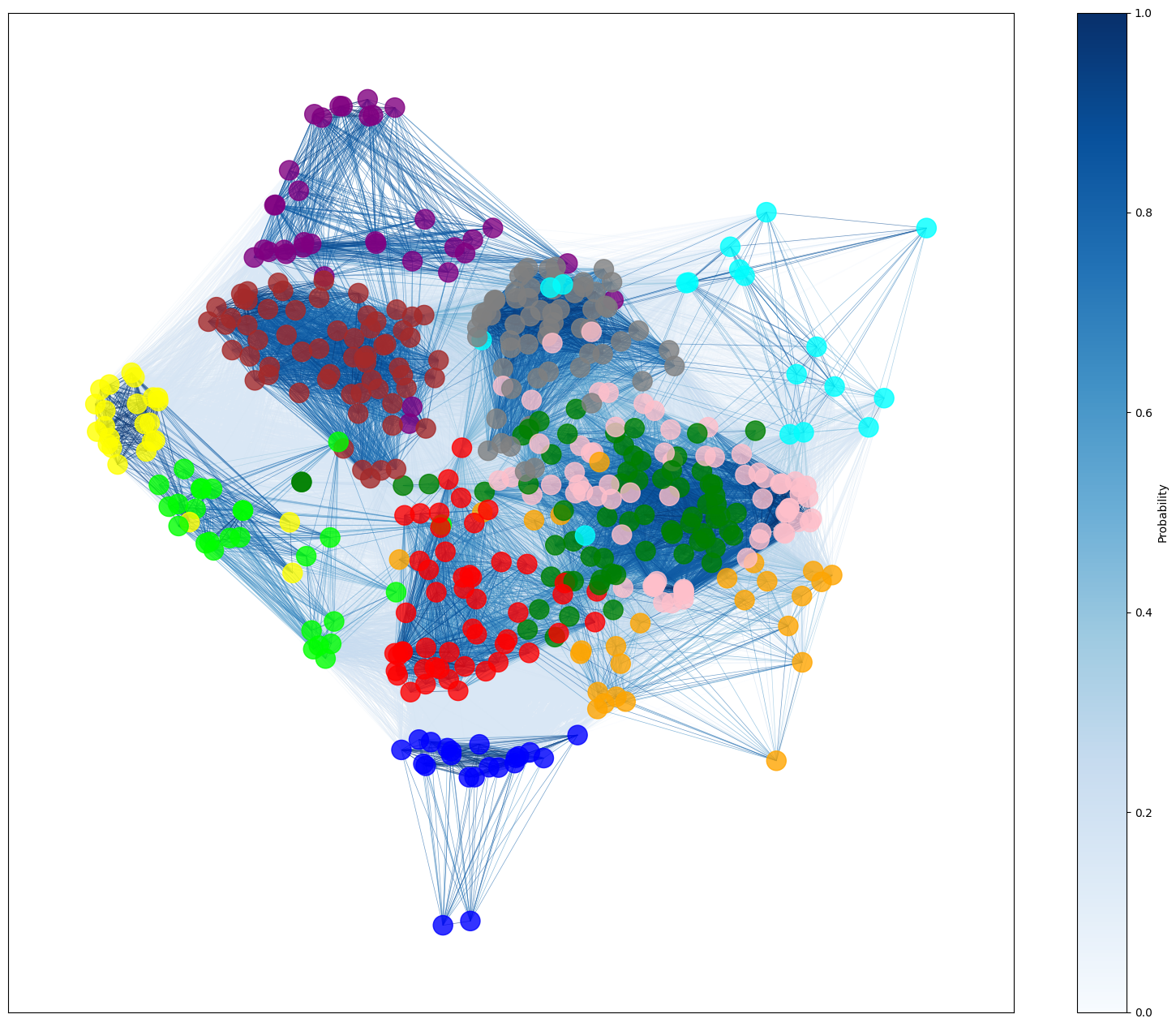}
        \caption{12-months look-back period}
        \label{fig:cluster_prob_12m}
    \end{subfigure}
    \hfill
    \begin{subfigure}[b]{0.45\textwidth}
        \centering
        \includegraphics[width=\textwidth]{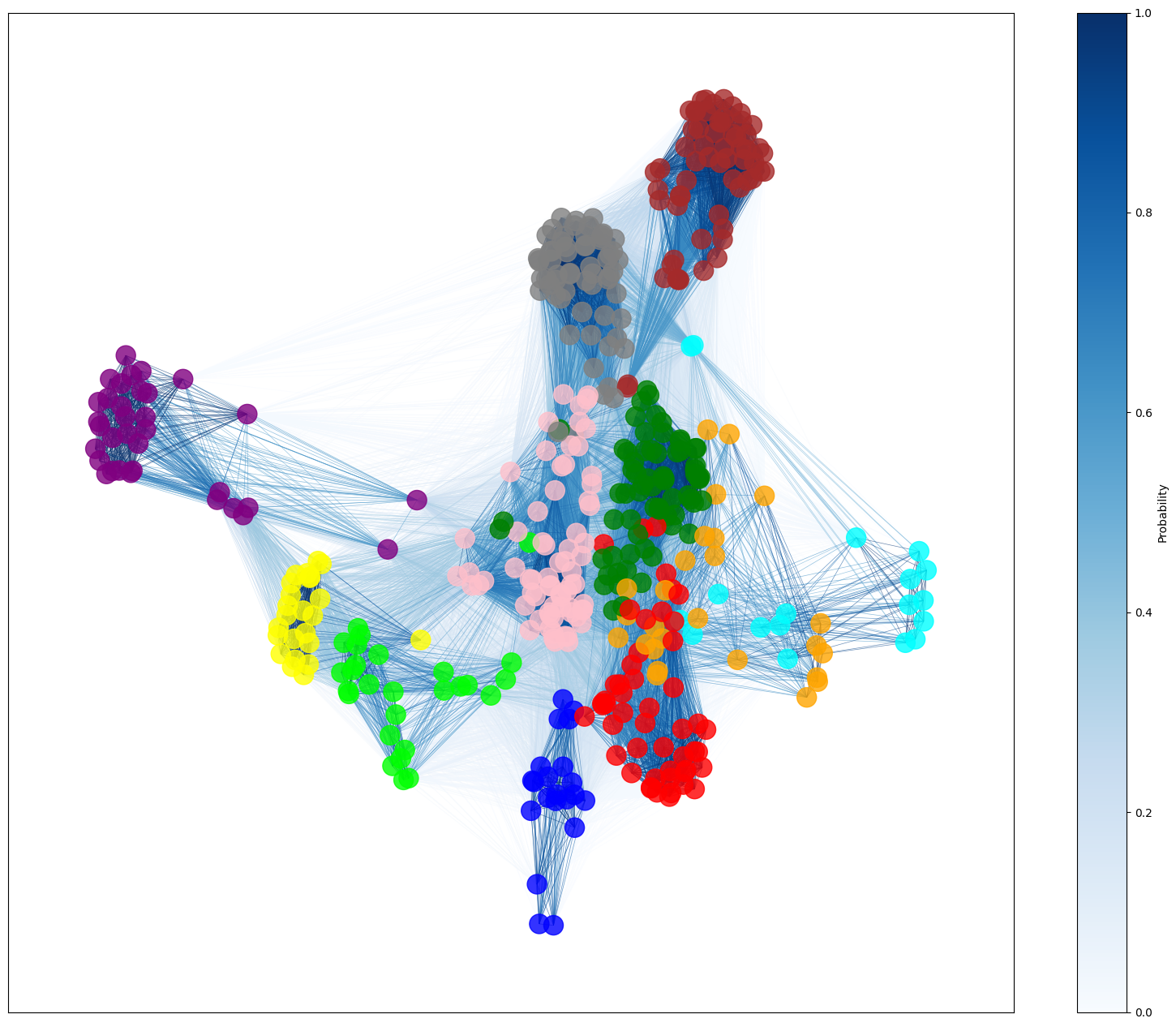}
        \caption{24-months look-back period}
        \label{fig:cluster_prob_24m}
    \end{subfigure}
    \caption{Probability of stocks being clustered together over different look-back periods.}
    \label{fig:cluster_prob}
\end{figure}

Figure \ref{fig:cluster_prob} shows four different networks based on the probability of stocks being clustered together. The darker the edge between two nodes, the more likely it is that these two stocks are clustered together. Each node is coloured based on the sector map in Table \ref{tab:sector_mapping}. Over every look-back, stocks within the energy sector are mostly clustered together. We generally see clear structures for most sectors except Communication Services. However, we can also note that in each period, there are several outliers. For example, in the 12-months look-back, we can clearly see two cyan dots within the grey Information Technology sector. These two dots represent the two share classes of Alphabet, which, regarding to the GICS classification, belong to Communication Services. However, due to the widespread business activities of Alphabet, intuitively it makes sense that this stock might be behaving similarly to Information Technology stocks. 

To better understand which stocks have high probabilities of being clustered with stocks from other sectors, we compute the weighted cross-sector probability and the most connected sector probability as follows. Let $S$ and $T$ denote two different stocks and Sectors be the set of all sectors. Then, $\text{sector}(S)$ and $\text{sector}(T)$ are the respective sector memberships of each stock. Lastly, $P(S,T)$ describes the probability of stocks $S$ and $T$ being connected. We can then define $S_{cross}$ as the total connection strength for all stocks $S$ with stocks from different sectors:

\begin{equation}
    S_{cross} = \sum_{\substack{S, T \in \text{stocks} \\ \text{sector}(T) \neq \text{sector}(S)}} P(S,T)
\end{equation}

$S_{same}$ can be defined as the total connection strength for all stocks $S$ with stocks from the same sector:

\begin{equation}
    S_{same} = \sum_{\substack{S, T \in \text{stocks} \\ \text{sector}(T) = \text{sector}(S)}} P(S,T)
\end{equation}

Combining these equations, we can compute $S_{total}$ which defines the total connection strength of all stocks with all other stocks:

\begin{equation}
    S_{total} = S_{cross} + S_{same}
\end{equation}

Lastly, we calculate $S_{most}$ which describes the total connection strength of all stocks with the most frequently connected sector that is different from their own sector:

\begin{equation}
    S_{most} = \max_{\substack{\text{sec} \in \text{Sectors} \\ \text{sec} \neq \text{sector}(S)}} \sum_{\substack{ S,T \in \text{stocks} \\ \text{sector}(T) = \text{sec}}} P(S,T)
\end{equation}

In Table \ref{tab:cluster_stats} we combine the equations above to compute the weighted cross-sector probability as

\begin{equation}
    P_{cross} = \frac{S_{cross}}{S_{total}}
\end{equation}

and the most connected sector probability as

\begin{equation}
    P_{most} = \frac{S_{most}}{S_{total}}
\end{equation}

For each period, we include the ten stocks with the highest $P_{cross}$.

\begin{table}[H]
\centering
\begin{adjustbox}{max width=\textwidth}
\begin{tabular}{llrrll}
\toprule
Period & Stock & \makecell{Weighted Cross-\\Sector Probability} & \makecell{Most Connected \\ Sector Probability} & Original Sector & Most Connected Sector \\
\midrule
\midrule
3 Months & Alphabet Inc. Class A & 0.9214 & 0.4447 & Communication Services & Information Technology \\
3 Months & Alphabet Inc. Class C & 0.9214 & 0.4447 & Communication Services & Information Technology \\
3 Months & Omnicom Group Inc. & 0.8907 & 0.3301 & Communication Services & Financials \\
3 Months & Interpublic Group of Companies Inc. & 0.8815 & 0.2957 & Communication Services & Financials \\
3 Months & Meta Platforms Inc. & 0.8765 & 0.3758 & Communication Services & Information Technology \\
3 Months & News Corp Class B & 0.8745 & 0.1984 & Communication Services & Financials \\
3 Months & News Corp Class A & 0.8708 & 0.1953 & Communication Services & Industrials \\
3 Months & Live Nation Entertainment Inc. & 0.8604 & 0.2481 & Communication Services & Consumer Discretionary \\
3 Months & Vulcan Materials Company & 0.8430 & 0.2754 & Materials & Industrials \\
3 Months & Ecolab Inc. & 0.8364 & 0.2111 & Materials & Industrials \\
\midrule
\midrule
6 Months & Alphabet Inc. Class A & 0.9129 & 0.5680 & Communication Services & Information Technology \\
6 Months & Alphabet Inc. Class C & 0.9129 & 0.5680 & Communication Services & Information Technology \\
6 Months & Visa Inc. & 0.8753 & 0.4646 & Financials & Information Technology \\
6 Months & Mastercard Inc. & 0.8749 & 0.4679 & Financials & Information Technology \\
6 Months & Eastman Chemical Company & 0.8259 & 0.3691 & Materials & Industrials \\
6 Months & LyondellBasell Industries N.V. & 0.8171 & 0.3923 & Materials & Industrials \\
6 Months & Celanese Corporation & 0.8023 & 0.3681 & Materials & Industrials \\
6 Months & Live Nation Entertainment Inc. & 0.8018 & 0.2241 & Communication Services & Consumer Discretionary \\
6 Months & Steel Dynamics Inc. & 0.7976 & 0.4195 & Materials & Industrials \\
6 Months & Vulcan Materials Company & 0.7944 & 0.2899 & Materials & Industrials \\
\midrule
\midrule
12 Months & Alphabet Inc. Class C & 0.9199 & 0.5331 & Communication Services & Information Technology \\
12 Months & Alphabet Inc. Class A & 0.9199 & 0.5331 & Communication Services & Information Technology \\
12 Months & Visa Inc. & 0.8998 & 0.4946 & Financials & Information Technology \\
12 Months & Mastercard Inc. & 0.8998 & 0.4907 & Financials & Information Technology \\
12 Months & LyondellBasell Industries N.V. & 0.8406 & 0.4626 & Materials & Industrials \\
12 Months & Eastman Chemical Company & 0.8319 & 0.4515 & Materials & Industrials \\
12 Months & Celanese Corporation & 0.8117 & 0.3581 & Materials & Industrials \\
12 Months & Steel Dynamics Inc. & 0.7980 & 0.3629 & Materials & Industrials \\
12 Months & Nucor Corporation & 0.7783 & 0.4663 & Materials & Industrials \\
12 Months & S\&P Global Inc. & 0.7704 & 0.3742 & Financials & Information Technology \\
\midrule
\midrule
24 Months & Eastman Chemical Company & 0.8465 & 0.4222 & Materials & Financials \\
24 Months & Alphabet Inc. Class C & 0.8164 & 0.5771 & Communication Services & Information Technology \\
24 Months & Alphabet Inc. Class A & 0.8164 & 0.5771 & Communication Services & Information Technology \\
24 Months & Interpublic Group of Companies Inc. & 0.8137 & 0.4139 & Communication Services & Financials \\
24 Months & Celanese Corporation & 0.7938 & 0.4294 & Materials & Financials \\
24 Months & LyondellBasell Industries N.V. & 0.7657 & 0.4516 & Materials & Financials \\
24 Months & Omnicom Group Inc. & 0.7651 & 0.3350 & Communication Services & Financials \\
24 Months & AT\&T Inc. & 0.7540 & 0.2546 & Communication Services & Financials \\
24 Months & Weyerhaeuser Company & 0.7187 & 0.3429 & Real Estate & Industrials \\
24 Months & DuPont de Nemours Inc. & 0.7123 & 0.2983 & Materials & Financials \\

\bottomrule
\end{tabular}
\end{adjustbox}
\caption{Probabilities of stocks being clustered with stocks from different sectors}
\label{tab:cluster_stats}
\end{table}

In Table \ref{tab:cluster_stats} we summarise the probabilities of stocks being clustered with stocks from different sectors and show, which sectors they are most often clustered with. For the 3-months look-back, the two share classes of Alphabet  exhibit the highest values for $P_{most}$. The most connected sector is Information Technology, which we have already seen in Figure \ref{fig:cluster_prob}. This behaviour is consistent over all periods. As mentioned above, intuitively this makes sense due to Alphabet's broad business activities and the technological nature of its operations.

Similarly, Omnicom Group and Interpublic Group show high values for the weighted cross-sector probability. Both these stocks are most often clustered with Financials. Interestingly, we can see both stocks in the 3-months and 24-months look-back, not in the medium-term look-backs. This could indicate that imminent and long-term trends are similar to Financials, while over the medium periods sector-own valuation factors or exogenous impacts could play a higher role. 

Meta Platforms also shows a high likelihood of cross-sector clustering, particularly with Information Technology, as reflected by its Most Connected Sector Probability of 0.38. This aligns with its significant technological and digital advertising operations.

In the 6-months look-back period, Visa and Mastercard join the ranking. Both of these firms are considered Financials, yet have high cross-sector probabilities. The most connected sector is Information Technology, which likely reflects their heavy reliance on technology for transaction processing and payments services.

For the 12-months look-back period, the clustering behaviour remains consistent. Alphabet, Visa, and Mastercard are showing similar patterns as they did in the shorter look-back periods. The Materials sector stocks such as LyondellBasell, Eastman Chemical and Celanese consistently show high cross-sector clustering with Industrials. As their business activities and products are heavily used in industrial applications, this connection makes sense, assuming that strong (or weak) performance of industrial corporates also benefits (or harms) the supply-chain companies. 

In the 24-months period we can see similar stocks and behaviour as before, which indicate robustness in our methodology as we continuously detect stocks where simple sector classification would not depict the actual market relationships. In this period, one notable addition is Weyerhaeuser , a Real Estate stock specialised on sustainable timber production, amongst other business segments. With a cross-sector probability of 0.72, WY shows notable connections to Industrials. This could be due to the industrial applications of its timber products.

Overall, the results indicate that certain stocks, particularly those with broad and diversified business operations, frequently cluster with sectors outside their original classification. Intuitively, this makes sense. It shows however, that relying on simple, pre-defined categories might not be sufficient in diversifying risk. This also supports our hypothesis four, showing that DynMSA is capable of finding a hidden market structure.

\subsubsection{Case study: COVID-19}

As exogenous shocks can have a significant impact on stock market structures, we had a closer look at the influences of the COVID-19 pandemic, particularly the change of community numbers over time during 2020. COVID-19 caused high volatility and changes in the correlation structure of the market. We wanted to see if our cluster ensemble approach was able to detect these changes and how the number of clusters changed over time, which is part of our third hypothesis. 

\begin{figure}
	\includegraphics[width = \textwidth]{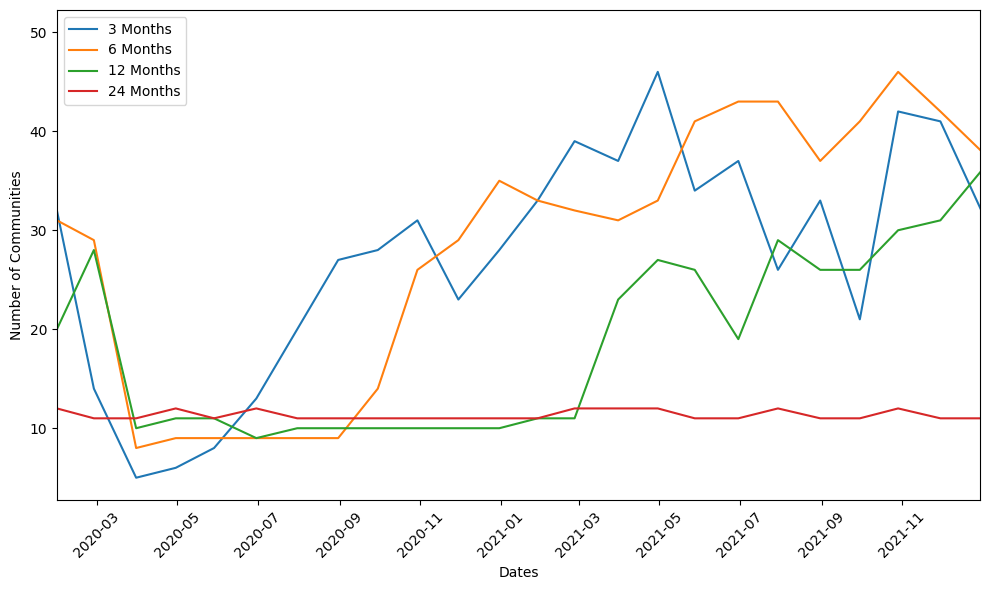}
	\caption{Number of clusters during 2020 and 2021 for S\&P 500}
	\label{fig:corona_spx}
\end{figure}

\begin{table}[ht]
\centering
	\begin{adjustbox}{max width=\textwidth}
        \begin{tabular}{lrrrrrrrrrrrr}
            \toprule
            & January & February & March & April & May & June & July & August & September & October & November & December \\
            \midrule
            Sector 3 Months & 0.1766 & 0.1260 & 0.0725 & 0.0794 & 0.0820 & 0.1123 & 0.1214 & 0.1376 & 0.1681 & 0.1414 & 0.1349 & 0.1459 \\
            Algorithm 3 Months & 0.4098 & 0.3138 & 0.0056 & 0.0577 & 0.1379 & 0.1944 & 0.3057 & 0.3360 & 0.3869 & 0.3566 & 0.3335 & 0.3947 \\
            Sector 24 Months & 0.1744 & 0.1720 & 0.1152 & 0.1139 & 0.1130 & 0.1098 & 0.1110 & 0.1118 & 0.1115 & 0.1087 & 0.1093 & 0.1101 \\
            Algorithm 24 Months & 0.1931 & 0.1725 & 0.0900 & 0.1322 & 0.0954 & 0.1282 & 0.0947 & 0.1008 & 0.1077 & 0.0967 & 0.1145 & 0.1103 \\
            \bottomrule
        \end{tabular}
	\end{adjustbox}
\caption{Difference of average intra- and inter-cluster correlation for S\&P 500 during 2020}
\label{tab:corona_spx_avg_corr}
\end{table}

Figure \ref{fig:corona_spx} and Table \ref{tab:corona_spx_avg_corr} show the number of clusters and the difference of average intra- and inter-cluster correlation for the S\&P 500 during 2020/2021 and 2020 respectively. We mainly focus our analysis on the two look-back horizons for our rolling correlation calculation of 3 and 24 months. We expected to see a heavier influence on the three months rolling correlations and indeed, we see that the number of clusters dropped significantly in April, containing the correlations from March 2020, the month many countries introduced lock downs. We also see strong fluctuations in the difference of average intra- and inter-cluster correlation during 2020. 

As we have noted above, we can relatively clearly determine when the month March leaves the correlation calculations as we exhibit a sudden increase around the 3-months mark. We can also see this for the 6 and 12 months. 

On the other side, using the 24 months look-back, correlations and number of cluster remain relatively stable. The number of clusters remains stable around 11-12 throughout the whole sample period. The differences in average intra- and inter-cluster correlations for the sector benchmark remain relatively consistent, moving around 0.11 throughout the year while fluctuating only lightly. Our DynMSA approach shows a slightly more pronounced initial drop in April, from 0.17 in February to 0.09 in March, but it stabilises quickly and remains within a narrow range for the rest of the year.

From a portfolio management and diversification perspective, the 24-months look-back period offers greater stability. While it may not fully capture the intricate and rapid changes in market correlations caused by an exogenous shock like the COVID-19 pandemic, it provides a balanced approach that maintains the integrity of the market's overall correlation structure. This stability is particularly valuable for long-term investment strategies, where sudden, short-term fluctuations can be mitigated by a more comprehensive view of market dynamics. 

\subsubsection{Performance comparison}
Lastly, we build portfolios out of our clusters to evaluate the practical implications of our clustering methodology. By constructing portfolios based on the top and bottom-performing stocks within each cluster, we aim to assess the potential returns and risks associated with our clustering approach. This section presents the performance of these portfolios over time and compares them with the overall market returns and aims to answer our last hypothesis about portfolio construction benefits of our methodology.

\begin{figure}
    \centering
    \begin{subfigure}[b]{0.45\textwidth}
        \centering
        \includegraphics[width=\textwidth]{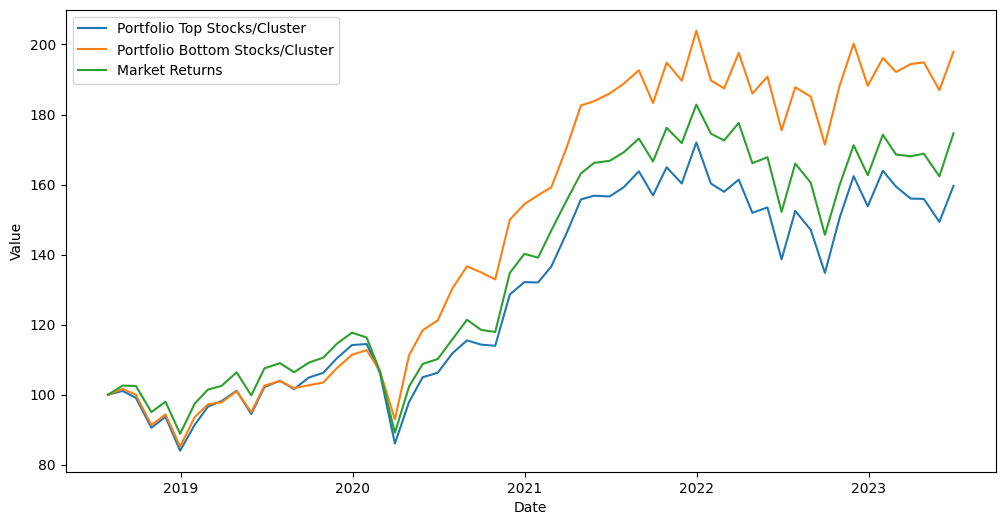}
        \caption{3-months look-back period}
        \label{fig:cluster_return_3m}
    \end{subfigure}
    \hfill
    \begin{subfigure}[b]{0.45\textwidth}
        \centering
        \includegraphics[width=\textwidth]{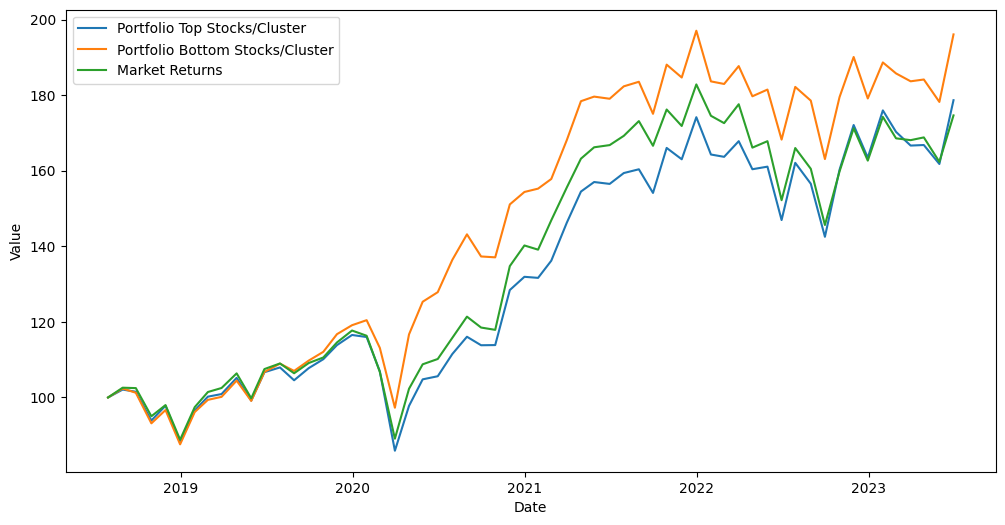}
        \caption{6-months look-back period}
        \label{fig:cluster_return_6m}
    \end{subfigure}
    
    \vspace{0.5cm} 
    
    \begin{subfigure}[b]{0.45\textwidth}
        \centering
        \includegraphics[width=\textwidth]{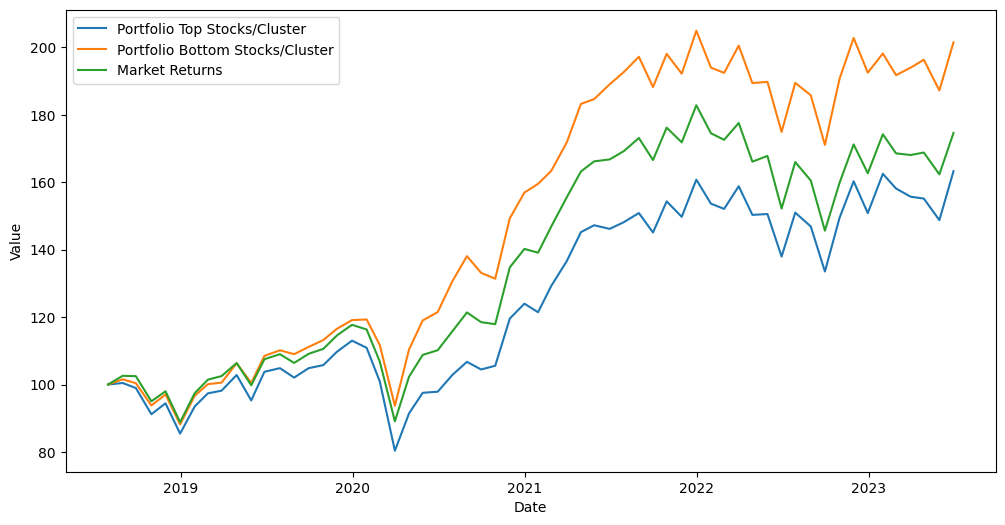}
        \caption{12-months look-back period}
        \label{fig:cluster_return_12m}
    \end{subfigure}
    \hfill
    \begin{subfigure}[b]{0.45\textwidth}
        \centering
        \includegraphics[width=\textwidth]{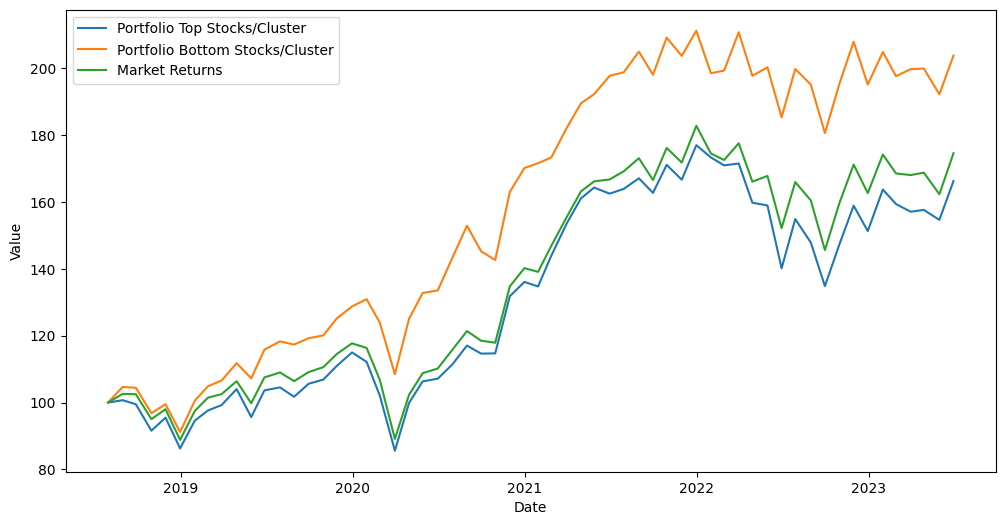}
        \caption{24-months look-back period}
        \label{fig:cluster_return_24m}
    \end{subfigure}
    \caption{Performance comparison of cluster portfolios vs an equally weighted market portfolio over different look-back periods.}
    \label{fig:cluster_returns}
\end{figure}

Figure \ref{fig:cluster_returns} shows the performance of two cluster-based portfolios against a market portfolio. The market portfolio invests equally in all stocks in the investment universe, rebalancing periodically. For the cluster portfolios, we used clusters from our DynMSA methodology as described in section \ref{sec:DynMSA}. Each portfolio consisted of $x=75$ equally weighted stocks, rebalanced monthly.

In the plot above we can see that the order of portfolio returns remains stable over different look-back periods. Using the bottom-ranked stocks from our clusters, in other words the stocks that least represent the cluster they are attributed to, always yield the highest returns. On the contrary, the top-ranked stocks underperform our market benchmark. This is an interesting result as one could assume, that the highest ranked stocks are the stocks that behave mostly contrary to the highest ranked stocks in the other clusters. We argue though, that considering diversification and cluster risk can explain these results. 

When constructing portfolios based on cluster analysis, the goal is to identify groups of stocks that exhibit similar behaviours or correlations. The top-ranked stocks within each cluster are those that exhibit the highest intra-cluster correlations, meaning they are the most representative of the cluster's overall behaviour. While this might seem beneficial, it also implies a higher concentration of cluster-specific risks. If a particular cluster experiences adverse conditions, the top-ranked stocks, being highly correlated with each other, are likely to suffer collectively, leading to significant losses in the portfolio. The bottom-ranked stocks on the other side exhibit much lower intra-cluster correlation and do not behave as synchronous as the top-ranked stocks do.  As a result, they are less susceptible to the specific risks associated with any single cluster. This inherent diversification within the bottom-ranked stocks makes them better candidates for reducing overall portfolio risk.

These aspects also align with the broader principle of diversification, which aims to spread investments across a wide range of uncorrelated assets to reduce risk. By selecting stocks that are not strongly correlated with each other, the bottom-ranked stocks inherently contribute to a more diversified and resilient portfolio. This reduced correlation means that the performance of the bottom-ranked stocks is less likely to be uniformly affected by sector-specific or market-wide events, leading to a smoother and potentially higher return profile over time.

To evaluate if the returns are smoother from a risk-adjusted perspective, we calculated key performance indicators (KPIs) for all portfolios, as shown in Table \ref{tab:combined_strategies}. The KPIs include annual return (Ann Ret), annual volatility (Ann Vola), Sharpe ratio, Sortino ratio, maximum drawdown (Max DD), downside volatility (Down Vola), beta, R-squared, and alpha. These metrics provide a comprehensive comparison of the cluster-based portfolios against the benchmark, highlighting the benefits of using our DynMSA methodology for portfolio construction.

\begin{table}[h]
\centering
\begin{adjustbox}{max width=\textwidth}
\begin{tabular}{lrrrrrrrrr}
\toprule
 & Ann Ret & Ann Vola & Sharpe  & Sortino  & Max DD & Down Vola & Beta & R-squared & Alpha \\
\midrule
\textbf{Strategy 3m} & & & & & & & & & \\
Benchmark & 0.1437 & 0.2066 & 0.6956 & 1.0419 & -0.2431 & 0.1379 & NaN & NaN & NaN \\
Portfolio Top Stocks/Cluster & 0.1251 & 0.2144 & 0.5833 & 0.8230 & -0.2482 & 0.1520 & 1.0237 & 0.9726 & NaN \\
Portfolio Bottom Stocks/Cluster & 0.1721 & 0.2037 & 0.8445 & 1.4925 & -0.1752 & 0.1153 & 0.9499 & 0.9279 & 0.0355 \\
\midrule
\textbf{Strategy 6m} & & & & & & & & & \\
Benchmark & 0.1437 & 0.2066 & 0.6956 & 1.0419 & -0.2431 & 0.1379 & NaN & NaN & NaN \\
Portfolio Top Stocks/Cluster & 0.1514 & 0.2158 & 0.7015 & 0.9884 & -0.2628 & 0.1532 & 1.0297 & 0.9715 & 0.0034 \\
Portfolio Bottom Stocks/Cluster & 0.1698 & 0.2037 & 0.8336 & 1.3694 & -0.1923 & 0.1240 & 0.9534 & 0.9349 & 0.0328 \\
\midrule
\textbf{Strategy 12m} & & & & & & & & & \\
Benchmark & 0.1437 & 0.2066 & 0.6956 & 1.0419 & -0.2431 & 0.1379 & NaN & NaN & NaN \\
Portfolio Top Stocks/Cluster & 0.1318 & 0.2204 & 0.5979 & 0.8661 & -0.2889 & 0.1521 & 1.0545 & 0.9773 & NaN \\
Portfolio Bottom Stocks/Cluster & 0.1771 & 0.2063 & 0.8585 & 1.4300 & -0.2148 & 0.1239 & 0.9755 & 0.9541 & 0.0369 \\
\midrule
\textbf{Strategy 24m} & & & & & & & & & \\
Benchmark & 0.1437 & 0.2066 & 0.6956 & 1.0419 & -0.2431 & 0.1379 & NaN & NaN & NaN \\
Portfolio Top Stocks/Cluster & 0.1349 & 0.2176 & 0.6199 & 0.9448 & -0.2558 & 0.1428 & 1.0394 & 0.9736 & NaN \\
Portfolio Bottom Stocks/Cluster & 0.1764 & 0.1914 & 0.9220 & 1.7167 & -0.1717 & 0.1028 & 0.8951 & 0.9338 & 0.0478 \\
\bottomrule
\end{tabular}
\end{adjustbox}
\caption{Combined Strategies Performance Metrics}
\label{tab:combined_strategies}
\end{table}

For the 3-months strategy, the Portfolio Bottom Stocks/Cluster significantly outperforms both the Portfolio Top Stocks/Cluster and the benchmark in terms of risk-adjusted returns. It achieves an annual return of 17.21\%, which surpasses the benchmark's 14.37\% and the Portfolio Top Stocks/Cluster's 12.51\%. The Sharpe and Sortino ratios for the Portfolio Bottom Stocks/Cluster are also superior, indicating better risk-adjusted performance. Notably, it has a lower maximum drawdown as well as a lower downside volatility than the other portfolios. This pattern remains stable over all periods, as indicated by the performance plots above. The risk-adjusted return metrics, mainly the Sortino ratio show that portfolios based upon our clustering method are constantly outperforming the benchmark, showing both better returns during market upsides but most importantly reduced downside risk, which is exactly what we wanted to achieve. The beta of the Portfolio Bottom Stocks/Cluster is also consistently lower, suggesting it is less volatile relative to the market. This is particularly true the longer the look-back periods are. While 12 months have the highest annualised return, 24 months have the highest risk-adjusted return as measured by the Sortino ratio due to the superior downside risk reduction.

Surprising but welcome was the fact that even this relatively naive portfolio selection process focused on risk mitigation and diversification yielded in some alpha. This highlights that our methodology was capable of detecting long-term trends in the markets particularly well and clustered accordingly.

\section{Conclusion}
In this research, we developed a novel, graph-based clustering algorithm called DynMSA, combining Random Matrix Theory with modularity optimisation and spectral clustering. This algorithm identifies clusters of stocks that are highly correlated within a cluster but un- or negatively correlated between clusters. We tested our methodology on the S\&P 500 universe and compared our results to both sector- and market-based benchmarks. Our contributions to the literature include:

\begin{itemize}
    \item Implementing a sector-based ground truth calibration for modularity optimisation.
    \item Using a correlation-based distance function inspired by the unweighted pair group method with arithmetic mean to compute the similarity matrix for spectral clustering instead of the standard Gaussian method.
\end{itemize}

Our primary objective was to find clusters and hidden market structures based on return correlations that offer inherent diversification to aid in risk-reducing portfolio allocation. To achieve this, we tested several hypotheses:

\begin{itemize}
    \item \textit{Intra- and inter-cluster correlation differences:} DynMSA significantly outperformed the baseline model in terms of intra- and inter-cluster correlation differences across different correlation look-back periods.
    \item \textit{Combined approach effectiveness:} The combined approach of modularity optimisation and spectral clustering improved key metrics, especially over medium-term correlation look-backs. However, for 24-months look-backs, spectral clustering did not improve these metrics, so we relied solely on modularity optimisation. Our improvements were statistically significant at the  p < 0.01  level, and the cluster selection was significantly different from random selection.
    \item \textit{Cluster stability and regime changes:} Our method identified stable clusters over time and detected regime changes due to exogenous shocks such as the COVID-19 pandemic. Post-pandemic, clusters returned to pre-pandemic levels. The Ukraine invasion did not significantly impact US market structures. We found several stocks, such as Alphabet, Meta, Visa, and Mastercard, frequently grouped outside their sector.
    \item \textit{Hidden market structure}: DynMSA was capable of finding stocks that are not perfectly attributable to certain sectors. When plotting the cluster strucute over time, a hidden market structure becomes clearly visible.
    \item \textit{Portfolio construction benefits:} Portfolios constructed using our clusters showed higher Sortino and Sharpe ratios, lower downside volatility, reduced maximum drawdown, and higher annualised returns compared to an equally weighted market benchmark. Notably, the best portfolios were based on stocks least correlated with their clusters, indicating additional diversification benefits.
\end{itemize}

We observed that longer look-back periods led to better portfolio performance, likely due to more stable correlations over time, making it easier for the algorithms to detect distinct patterns.

Future work could explore the dynamics of stocks typically clustered together and identify periods of divergence and convergence to capture arbitrage opportunities. Additionally, incorporating macroeconomic and firm-specific indicators into our clustering ensemble could enhance cluster accuracy through multiplex graph analysis. Lastly it would be interesting to use our risk-based clusters and add an alpha-seeking component to the algorithm.

\newpage

\bibliographystyle{unsrtnat}
\bibliography{literature}

\end{document}